\documentclass[preprint,12pt]{elsarticle}

\usepackage{amssymb}

\biboptions{sort&compress}

\newlength{\dslashwidth}

\newcommand{\bsg}{\ensuremath{b\to X_s\gamma}}

\newcommand{\tb}{\ensuremath{\tan\beta}}

\newcommand{\beq}{\begin{equation}} 
\newcommand{\eeq}{\end{equation}}
\newcommand{\beqa}{\begin{eqnarray}} 
\newcommand{\eeqa}{\end{eqnarray}}
\newcommand{\newc}{\newcommand}
\newcommand{\bq}{\begin{equation}}
\newcommand{\eq}{\end{equation}}
\newcommand{\ba}{\begin{array}}
\newcommand{\ea}{\end{array}}
\newcommand{\bqa}{\begin{eqnarray}}
\newcommand{\eqa}{\end{eqnarray}}

\newcommand{\lnf}{{\ifmmode \Lambda^{(N_f)} \else $\Lambda^{(N_f)}$\fi}}
\newcommand{\ms}{{\ifmmode \overline{MS} \else $\overline{MS}$\fi}}
\newcommand{\dr}{{\ifmmode \overline{DR} \else $\overline{DR}$\fi}}
\newcommand{\lms}{{\ifmmode \Lambda^{(5)}_{\overline{MS}} \else $\Lambda^{(5)}_{\overline{MS}}$\fi}}
\newcommand{\lam}{{\ifmmode \Lambda \else $\Lambda$\fi}}
\newcommand{\mev}{{\ifmmode {\rm MeV} \else ${\rm MeV}$\fi}}
\newcommand{\gev}{{\ifmmode {\rm GeV} \else ${\rm GeV}$\fi}}
\newcommand{\gevc}{{\ifmmode {\rm GeV/c^2} \else ${\rm GeV/c^2}$\fi}}
\newcommand{\tev}{{\ifmmode {\rm TeV} \else ${\rm TeV}$\fi}}
\newcommand{\tevc}{{\ifmmode {\rm TeV/c^2} \else ${\rm TeV/c^2}$\fi}}
\newcommand{\lp}{{\ifmmode L^+  \else $L^+$\fi}}
\newcommand{\lm}{{\ifmmode L^-  \else $L^-$\fi}}
\newcommand{\mlp}{{\ifmmode M(L^-) \else $M(L^-)$\fi}}
\newcommand{\mlz}{{\ifmmode M(L^0) \else $M(L^0)$\fi}}
\newcommand{\lz}{{\ifmmode L^0 \else $L^0$\fi}}
\newcommand{\ev}{{\ifmmode GeV/c^2 \else $GeV/c^2$\fi}}
\newcommand{\tri}{{\ifmmode \triangleup \else $\triangleup$\fi}}
\newcommand{\unl}{{\ifmmode U_{lL^0} \else $U_{lL^0}$\fi}}\newcommand{\gL}{{\ifmmode g_L \else $g_{L}$\fi}}
\newcommand{\gR}{{\ifmmode g_R  \else $g_{R}$\fi}}
\newcommand{\gumu}{{\ifmmode \gamma^{\mu} \else $\gamma^{\mu}$\fi}}
\newcommand{\gunu}{{\ifmmode \gamma^{\nu} \else $\gamma^{\nu}$\fi}}
\newcommand{\gdmu}{{\ifmmode \gamma_{\mu} \else $\gamma_{\mu}$\fi}}
\newcommand{\gdnu}{{\ifmmode \gamma_{\nu} \else $\gamma_{\nu}$\fi}}
\newcommand{\stw}{{\ifmmode\sin^2\theta_W \else $\sin^{2}\theta_{W}$ \fi}}
\newcommand{\sws}{{\ifmmode \;\sin^2\theta_W  \else $\;\sin^{2}\theta_{W}$ \fi}}
\newcommand{\cws}{{\ifmmode \;\cos^2\theta_W  \else $\;\cos^{2}\theta_{W}$ \fi}}
\newcommand{\sw}{{\ifmmode \;\sin\theta_W  \else $\sin\theta_{W}$ \fi}}
\newcommand{\cw}{{\ifmmode \;\cos\theta_W  \else $\;\cos\theta_{W}$ \fi}}
\newcommand{\tw}{{\ifmmode \;\tan\theta_W  \else $\;\tan\theta_{W}$ \fi}}
\newcommand{\qq}{{\ifmmode q\overline{q} \else $q\overline{q}$\fi}}
\newcommand{\lR}{{\ifmmode l_R \else $l_R$\fi}}
\newcommand{\lL}{{\ifmmode l_L \else $l_L$\fi}}
\newcommand{\nt}{{\ifmmode \nu_{\tau} \else $\nu_{\tau}$\fi}}
\newcommand{\nuR}{{\ifmmode \nu_R  \else $\nu_R$\fi}}
\newcommand{\nuL}{{\ifmmode \nu_L  \else $\nu_L$\fi}}
\newcommand{\qR}{{\ifmmode g_R  \else $q_R$\fi}}
\newcommand{\qL}{{\ifmmode q_L  \else $q_L$\fi}}
\newcommand{\qRp}{{\ifmmode q_R'  \else $q_{R}$'\fi}}
\newcommand{\qLp}{{\ifmmode q_L'  \else $q_{L}$'\fi}}
\newcommand{\est}{{\ifmmode e^{\bf \ast} \else $e^{\bf \ast}$\fi}}
\newcommand{\lst}{{\ifmmode l^{\bf \ast} \else $l^{\bf \ast}$\fi}}
\newcommand{\must}{{\ifmmode \mu^{\bf \ast} \else $\mu^{\bf \ast}$\fi}}
\newcommand{\taust}{{\ifmmode \tau^{\bf \ast} \else $\tau^{\bf \ast}$ \fi}}
\newcommand{\pperp}{{\ifmmode p_t  \else $p_t$\fi}}
\newcommand{\et}{{\ifmmode E_t  \else $E_t$\fi}}
\newcommand{\xt}{{\ifmmode x_t  \else $x_t$\fi}}
\newcommand{\smumu}{{\ifmmode \sigma_{\mu\mu}  \else $\sigma_{\mu\mu}$ \fi}}
\newcommand{\eg}{{\ifmmode e\gamma  \else $e\gamma$\fi}}
\newcommand{\epem}{{\ifmmode e^+e^-  \else $e^+e^-$\fi}}
\newcommand{\lplm}{{\ifmmode L^+L^-  \else $L^+L^-$\fi}}
\newcommand{\pp}{{\ifmmode p\overline p  \else $p\overline p$\fi}}
\newcommand{\llz}{{\ifmmode L^0\overline{L}^0 \else $L^0\overline{L}^0$\fi}}
\newcommand{\epemt}{{\ifmmode e^+e^- \to  \else $e^+e^- \to$\fi}}
\newcommand{\eb}{{\ifmmode E_{beam}  \else $E_{beam}$\fi}}
\newcommand{\ip}{{\ifmmode pb^{-1}  \else $pb^{-1}$\fi}}
\newcommand{\upm}{{\ifmmode ^{\pm}  \else $^{\pm}$\fi}}
\newcommand{\de}{{\ifmmode ^{\circ}  \else $^{\circ}$ \fi}}
\newcommand{\appr}{{\ifmmode \sim \else $\sim$ \fi}}
\newcommand{\corresp}{{\ifmmode \stackrel{\wedge}{=} \else $\stackrel{\wedge}{=}$ \fi}}
\newcommand{\sqrts}{{\ifmmode \sqrt{s} \else $\sqrt{s}$\fi}}
\newcommand{\zz}{{\ifmmode Z^0  \else $Z^0$\fi}}
\newcommand{\mz}{{\ifmmode M_{Z}  \else $M_{Z}$\fi}}
\newcommand{\mzs}{{\ifmmode M_{Z}^2  \else $M_{Z}^2$\fi}}
\newcommand{\mw}{{\ifmmode M_{W}  \else $M_{W}$\fi}}
\newcommand{\mws}{{\ifmmode M_{W}^2  \else $M_{W}^2$\fi}}
\newcommand{\mh}{{\ifmmode M_{Higgs}  \else $M_{Higgs}$\fi}}
\newcommand{\gt}{{\ifmmode \Gamma_{tot} \else $\Gamma_{tot}$\fi}}
\newcommand{\msusy}{{\ifmmode M_{SUSY}  \else $M_{SUSY}$\fi}}
\newcommand{\msusys}{{\ifmmode M_{SUSY}^2  \else $M_{SUSY}^2$\fi}}
\newcommand{\su}{{\ifmmode SU(3)_C\otimes\- SU(2)_L\otimes\- U(1)_Y \else $SU(3)_C\otimes SU(2)_L\otimes U(1)_Y$\fi}}
\newcommand{\suthree}{{\ifmmode SU(3)_C  \else $SU(3)_C$\fi}}
\newcommand{\sutwo}{{\ifmmode  SU(2)_L\otimes U(1)_Y \else $SU(2)_L\otimes U(1)_Y$\fi}}
\newcommand{\taup}{{\ifmmode \tau_{proton} \else $\tau_{proton}$\fi}}
\newcommand{\as}{{\ifmmode \alpha_{s}  \else $\alpha_{s}$\fi}}
\newcommand{\mgut}{{\ifmmode M_{GUT}  \else $M_{GUT}$\fi}}
\newcommand{\mguts}{{\ifmmode M_{GUT}^2  \else $M_{GUT}^2$\fi}}
\newcommand{\mzero}{{\ifmmode m_0        \else $m_0$\fi}}
\newcommand{\mhalf}{{\ifmmode m_{1/2}    \else $m_{1/2}$\fi}}
\newcommand{\sq}{{\ifmmode \tilde{q}    \else $\tilde{q}$\fi}}
\newcommand{\gl}{{\ifmmode \tilde{g}    \else $\tilde{g}$\fi}}
\newcommand{\mb}{{\ifmmode m_{b}    \else $m_{b}$\fi}}
\newcommand{\mt}{{\ifmmode m_{t}    \else $m_{t}$\fi}}
\newcommand{\mts}{{\ifmmode m_{t}^2    \else $m_{t}^2$\fi}}

\newcommand{\mtau}{{\ifmmode m_{\tau}  \else $m_{\tau}$\fi}}
\newcommand{\dpp}{{\ifmmode \delta_{pert} \else $\delta_{pert}$\fi}}
\newcommand{\dnp}{{\ifmmode\delta_{non-pert}\else$\delta_{non-pert}$\fi}}
\newcommand{\dew}{{\ifmmode \delta_{\rm EW}\else $\delta_{\rm EW}$\fi}}
\newcommand{\rt}{{\ifmmode R_{\tau}  \else $R_{\tau} $\fi}}
\newcommand{\rz}{{\ifmmode R_{Z}  \else $R_{Z} $\fi}}

\newcommand{\swb}{{\ifmmode \sin^2\theta_{\overline{MS}} \else $\sin^2\theta_{\overline{MS}}$\fi}}
\newcommand{\cwb}{{\ifmmode \cos^2\theta_{\overline{MS}} \else $\cos^2\theta_{\overline{MS}}$\fi}}

\newcommand{\bsmm}{\ensuremath{B^0_s\to\mu^+\mu^-}}

\newc\AIPCP[3] {{\em AIP Conf. Proc.} {\bf #1} (#2) #3}
\newc\AJ[3] {{\em Astrophys. J.} {\bf #1} (#2) #3}
\newc\AMS[3] {{\em Ann. Math. Statist.} {\bf #1} (#2) #3}
\newc\AP[3] {{\em Ann. Phys.} {\bf #1} (#2) #3}
\newc\APJ[3] {{\em Astropart. J.} {\bf #1} (#2) #3}
\newc\APP[3] {{\em Astropart. Phys.} {\bf #1} (#2) #3}
\newc\APS[3] {{\em Astrophys. J. Suppl.} {\bf #1} (#2) #3}
\newc\ARNPS[3] {{\em Ann. Rev. Nucl. Part. Sci.} {\bf C#1} (#2) #3}
\newc\BA[3] {{\em Bayesian Anal.} {\bf C#1} (#2) #3}
\newc\CPC[3] {{\em Comput. Phys. Commun.} {\bf C#1} (#2) #3}
\newc\CP[3] {{\em Contemp. Phys.} {\bf #1} (#2) #3}
\newc\EPJ[3] {{\em Euro. Phys. Journ.} {\bf C#1} (#2) #3}
\newc\JCAP[3] {{\em JCAP} {\bf #1} (#2) #3}
\newc\JHEP[3] {{\em JHEP} {\bf #1} (#2) #3}
\newc\JPG[3] {{\em J. Phys.} {\bf G #1} (#2) #3}
\newc\IJMP[3] {{\em Int. J. Mod. Phys.} {\bf A #1} (#2) #3}
\newc\MNRAS[3] {{\em Mon. Not. Roy. Astron. Soc.} {\bf #1} (#2) #3}
\newc\MPL[3] {{\em Mod. Phys. Lett.} {\bf A #1} (#2) #3}
\newc\NAR[3] {{\em New Astron. Rev.} {\bf #1} (#2) #3}
\newc\NCA[3] {{\em Nuovo Cimento} {\bf #1} (#2) #3}
\newc\NIM[3] {{\em Nucl. Instrum. Methods} {\bf #1} (#2) #3}
\newc\NIMA[3] {{\em Nucl. Instrum. Methods} {\bf A #1} (#2) #3}
\newc\NAT[3] {{\em Nature} {\bf #1} (#2) #3}
\newc\NPB[3] {{\em Nucl. Phys.} {\bf B #1} (#2) #3}
\newc\NPA[3] {{\em Nucl. Phys.} {\bf A #1} (#2) #3}
\newc\NPPS[3] {{\em Nucl. Phys. Proc. Suppl.} {\bf #1} (#2) #3}
\newc\PLB[3] {{\em Phys. Lett.} {\bf B #1} (#2) #3}
\newc\PR[3] {{\em Phys. Rep.} {\bf #1} (#2) #3}
\newc\PRL[3] {{\em Phys. Rev. Lett.} {\bf #1} (#2) #3}
\newc\PRD[3] {{\em Phys. Rev.} {\bf D #1} (#2) #3}
\newc\PRC[3] {{\em Phys. Rev.} {\bf C #1} (#2) #3}
\newc\PTP[3] {{\em Prog. Theor. Phys.} {\bf #1} (#2) #3}
\newc\RMP[3] {{\em Rev. Mod. Phys.} {\bf #1} (#2) #3 }
\newc\RPP[3] {{\em Rept. Prog. Phys.} {\bf #1} (#2) #3 }
\newc\SC[3] {{\em Science} {\bf #1} (#2) #3 }
\newc\ZPC[3] {{\em Z. Phys.} {\bf C #1} (#2) #3}
\newc\Err[3] {{\em Erratum-ibid.} {\bf #1} (#2) #3 }

\journal{Physics Letters B}

\begin{document}

\begin{frontmatter}


%
\title{A comparison of the Higgs sectors of the  CMSSM and NMSSM  for a 126 GeV Higgs boson}

\author[label1]{C. Beskidt}\ead{conny.beskidt@kit.edu}
\author[label1]{W. de Boer}\ead{wim.de.boer@kit.edu}
\author[label1,label2]{D.I. Kazakov}
\address[label1]{Institut f\"ur Experimentelle Kernphysik, Karlsruhe Institute of Technology, P.O. Box 6980, 76128 
Karlsruhe, Germany}
\address[label2]{Bogoliubov Laboratory of Theoretical Physics, Joint Institute for Nuclear Research, 141980, 6 
Joliot-Curie, Dubna, Moscow Region, Russia}

\begin{abstract}

The recent discovery of a Higgs-like boson at the LHC  with a mass of 126 GeV has revived the interest in supersymmetric 
models, which predicted a Higgs boson mass below 130 GeV long before its discovery.
We compare systematically the allowed parameter space in the constrained Minimal Supersymmetric Standard Model (CMSSM) 
and the Next-to-Minimal Supersymmetric Model (NMSSM) by minimizing the $\chi^2$ function 
with respect to all known constraints from accelerators and cosmology using GUT scale parameters. For the CMSSM the 
Higgs boson mass at tree level is below the $Z^0$ boson mass and large radiative corrections are needed to obtain a 
Higgs boson mass of 126 GeV, which requires stop squark masses  in the multi-TeV range. In contrast, for the NMSSM  
light stop quarks are allowed, since in the NMSSM at tree level the Higgs boson mass can be above the $Z^0$ boson mass 
from mixing with the additional singlet Higgs boson. 
Predictions for the scalar boson masses are given in both models with emphasis on the unique signatures of the NMSSM, 
where the heaviest scalar Higgs boson decays in the two lighter scalar Higgs bosons with a significant branching ratio, 
in which case one should observe double Higgs boson production at the LHC. Such a signal is strongly suppressed in the 
CMSSM. In addition, since the LSP is higgsino-like, Higgs boson decays into LSPs can be appreciable, thus leading to 
invisible Higgs decays.
\end{abstract}

\end{frontmatter}


\section{Introduction}
\label{Introduction}

The discovery of a 126 GeV Higgs boson \cite{Aad:2012tfa,Chatrchyan:2012ufa}  is  exciting for several reasons:
i) Its branching ratios to the electroweak gauge bosons and hence its interactions with them are in agreement with the 
prediction from the Higgs mechanism, that this interaction is proportional to the gauge boson masses.
ii) In the SM electroweak symmetry breaking (EWSB) is introduced ad hoc and the Higgs boson mass is not predicted, so it 
could have  any value between the electroweak scale and the TeV scale.  However, in the supersymmetric extension of the 
SM (SUSY) EWSB  {\it is } predicted (by radiative corrections) and the mass of the lightest Higgs boson is predicted to 
be below 130 GeV, as observed.
This  may be the strongest hint for SUSY in spite of the fact that the predicted supersymmetric particles  have not been 
discovered so far.

Additional hints for SUSY are the unification of gauge and Yukawa couplings at a large scale, the GUT scale as expected 
in Grand Unified Theories, the absence of quadratic divergencies to the Higgs boson mass, and the prediction of a dark 
matter particle with a correct relic density, see reviews, e.g. 
\cite{Haber:1984rc,deBoer:1994dg,Jungman:1995df,Martin:1997ns,Kazakov:2010qn}.
However, also a few problems exist. First of all, at tree level the Higgs boson mass is predicted to be below the 
$Z^0$-mass in the CMSSM  and to obtain a mass of 126 GeV requires large radiative corrections from stop loops, which are 
required to be in the multi-TeV range and/or fine-tuned maximal mixing in the stop sector.

Alternatively, one can consider the NMSSM, which has attracted much attention in the last year 
\cite{Cao:2013si,Cerdeno:2013cz,Kang:2013rj,Gherghetta:2012gb,King:2012tr,Kowalska:2012gs,Choi:2012he,Gogoladze:2012jp,
Agashe:2012zq,Bae:2012am,Gunion:2012gc,Graf:2012hh,Rathsman:2012dp,Jeong:2012ma,Almarashi:2012ri,Ellwanger:2012ke,
Vasquez:2012hn,Cao:2012fz,King:2012is,Gunion:2012zd,Ellwanger:2011aa,Ender:2011qh,Almarashi:2011qq,Stal:2011cz,
Almarashi:2011te,Almarashi:2011bf,Cao:2011re,Almarashi:2011hj,Mahmoudi:2010xp,Ananthanarayan:2013fga,Badziak:2013bda}, since it has additional contributions at 
tree level from the mixing with a Higgs singlet, so it does not need large radiative corrections from heavy stop quarks. 
The Higgs singlet was introduced long before in many models and it provides a solution to  the so-called $\mu$-problem, 
see e.g. Refs. \cite{Miller:2003ay,Ellwanger:2009dp}. 

In this Letter we compare the predictions of the CMSSM and NMSSM in the whole parameter space by optimizing the GUT 
scale parameters (4 in the CMSSM, 9 in the NMSSM) using all constraints from accelerators and cosmology (assuming the 
LSP forms the dark matter).
Studies in a similar direction \cite{Kowalska:2012gs} considered less free parameters implying more tight GUT scale 
relations in the Higgs sector. Relaxing these constraints allows for a significantly larger region of parameter space, 
especially the specific NMSSM radiative corrections \cite{Degrassi:2009yq}, which lower the Higgs boson mass by up to 3 
GeV, can be applied without a severe restriction on the parameter space. 
 
 Another comparison has been  done in Ref. \cite{Cao:2012fz}, albeit without GUT scale relations in that case, so the 
 masses are defined at the TeV scale. 
We study the restricted parameter space from GUT scale scenarios, since these allow to use the full 
radiative corrections  to SUSY masses and couplings.
In case the  SUSY masses are defined at a low scale, the radiative corrections are only integrated between the low scale and the  actual mass, which in practice means the masses are close to the tree level masses, thus effectively eliminating radiative corrections.
 In addition, with low scale definitions one ignores the fixed point solutions of the couplings, see Ref. \cite{Yeghian:1998km} and references therein.  These fixed point solutions do not allow maximal mixing in the stop sector, which implies a low value for the lightest Higgs, thus requiring multi-TeV stop masses in the CMSSM for a 126 GeV Higgs boson.   Still large differences are found in the restricted parameter space from GUT scale defined scenarios between the CMSSM and NMSSM, which are the key  to distinguish between the models in case more Higgs bosons will 
be found.

\section{The NMSSM}

 The Higgs fields of
the NMSSM consist of the usual two Higgs doublets together with
an additional complex Higgs  singlet S:
\begin{equation}  
H_u= \left( \begin{array}{c} H_u^+ \\ H_u^0 \end{array} \right), 
\quad H_d= \left( \begin{array}{c} H_d^0 \\ H_d^- \end{array} \right),
\quad S. 
\end{equation}

The additional parameters in the NMSSM originate from the additional couplings with the singlet. These additional terms 
are usually written as \cite{Miller:2003ay}:
\begin{eqnarray} \label{NMSSM}
 W_{\rm NMSSM}&=&W_F + \lambda\hat{H_u} \cdot \hat{H_d} \hat{S}
 + \frac{1}{3}\kappa \hat{S^3},\\
V_{\rm soft}^{\rm NMSSM}&=&\tilde m_u^2|H_u|^2 + \tilde m_d^2|H_d|^2
+ \tilde m_S^2|S|^2 +(A_\lambda \lambda SH_u\cdot H_d
+\frac{A_\kappa}{3}\kappa S^3).
\end{eqnarray}
Here $W_F$ is the superpotential of the MSSM without the $\mu$ term,
the dimensionless parameters $\lambda$ and $\kappa$ are the coefficients
 of the Higgs self couplings, and $\tilde{m}_{u}$, $\tilde{m}_{d}$, $\tilde{m}_{S}$,
$A_\lambda$ and $A_\kappa$ are the soft-breaking parameters.

After electroweak symmetry breaking the three soft breaking masses squared for
$H_u$, $H_d$ and $S$ can be expressed in terms of their vevs (i.e. $v_u$, $v_d$ and $s$)
through the minimization conditions of the scalar potential.
We assume all parameters to be real for the present study. In addition, we have the standard GUT scale parameters of the 
CMSSM, i.e. $\mzero$, $\mhalf$ and the trilinear couplings $A_t$, $A_b$ and $A_\tau$. The latter ones are taken to be 
unified at the GUT scale to $A_0$, so in total we have nine free parameters: 
\begin{equation}
 \mzero,~ \mhalf,~ \tan\beta, ~A_0,  ~ \lambda, ~\kappa,  ~A_\lambda, ~A_\kappa, ~\mu_{eff}\equiv\lambda s .
\label{params}
\end{equation}
In the CP-conserving NMSSM,
the neutral components from the three Higgs doublets ($S_i$) mix to form three physical CP-even
Higgs bosons, and $P_1$ and  $P_2$ mix to form two physical CP-odd
Higgs bosons. 

Under the basis ($S_1$, $S_2$, $S_3$), the elements of the mass matrix for $S_i$ fields at
tree level are given by \cite{Miller:2003ay}:
\begin{eqnarray}
{\cal M}^2_{11}&=&M_A^2+(m_Z^2-\lambda^2v^2)\sin^22\beta,\nonumber\\
{\cal M}^2_{12}&=&-\frac{1}{2}(m_Z^2-\lambda^2v^2)\sin4\beta,\nonumber\\
{\cal M}^2_{13}&=&-\frac{1}{2}(M_A^2\sin2\beta+\frac{2\kappa\mu^2}{\lambda})\frac{\lambda v}{\mu}\cos2\beta,\nonumber\\
{\cal M}^2_{22}&=&m_Z^2\cos^22\beta+\lambda^2v^2\sin^22\beta,\label{mix}\\
{\cal M}^2_{23}&=& 2 \lambda \mu v \left[1 - (\frac{M_A \sin 2\beta}{2 \mu} )^2
-\frac{\kappa}{2 \lambda}\sin2\beta\right],\nonumber\\
{\cal M}^2_{33}&=& \frac{1}{4} \lambda^2 v^2 (\frac{M_A \sin 2\beta}{\mu})^2
+ \frac{\kappa\mu}{\lambda} (A_\kappa +  \frac{4\kappa\mu}{\lambda} )
 - \frac{1}{2} \lambda \kappa v^2 \sin 2 \beta,\nonumber 
\end{eqnarray}
where ${\cal M}^2_{22}$ is nothing but $m_h^2$ at tree level without considering the mixing
among $S_i$, and its second term $\lambda^2v^2\sin^2 2\beta$ originates from the coupling
\mbox{$\lambda\hat{H_u} \cdot \hat{H_d} \hat{S}$} in Eq. \ref{NMSSM}. In the mixing matrix 
 \begin{equation}
M_A^2 \equiv \frac{2 \mu (A_\lambda + \kappa s)}{\sin 2 \beta},
\label{MA}
\end{equation}
   which is called $M_A$, since in the expression for $M^2_{11}$ it corresponds to the pseudo-scalar Higgs boson mass in 
   the MSSM limit of small $\lambda$. 
   $M^2_{33}$ corresponds to the diagonal term for the additional scalar  Higgs boson not present in the MSSM. In 
   addition to this additional scalar Higgs boson, the NMSSM has also an additional pseudo-scalar Higgs boson, which is 
   usually lighter and denoted by $A_1$. The mass of $A_2$ is usually close to the heaviest scalar Higgs boson, as in 
   the MSSM.
 \begin{figure} 
 \begin{center}
 \includegraphics[width=0.49\textwidth]{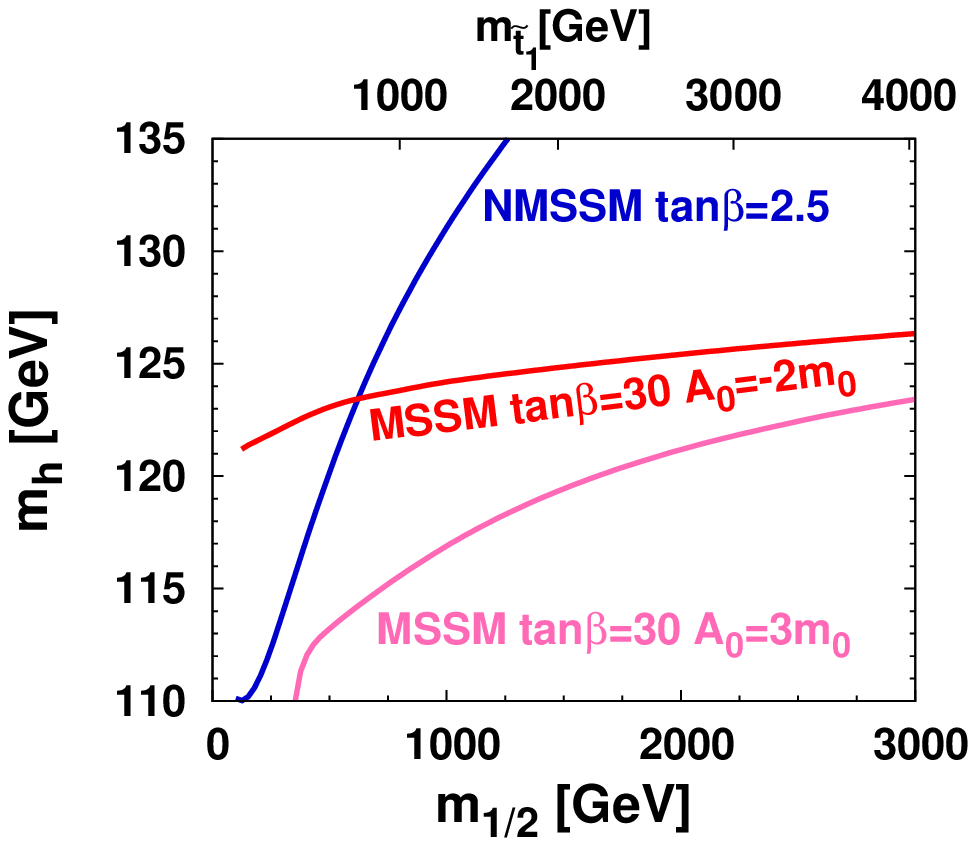}
 \includegraphics[width=0.49\textwidth]{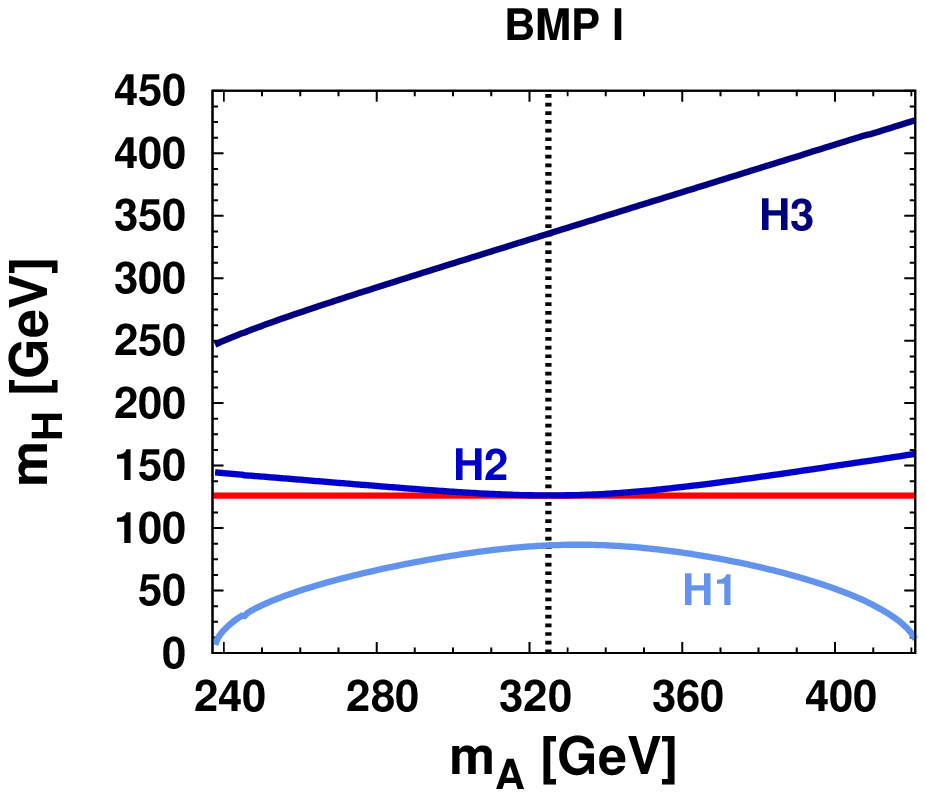} 
 \includegraphics[width=0.49\textwidth]{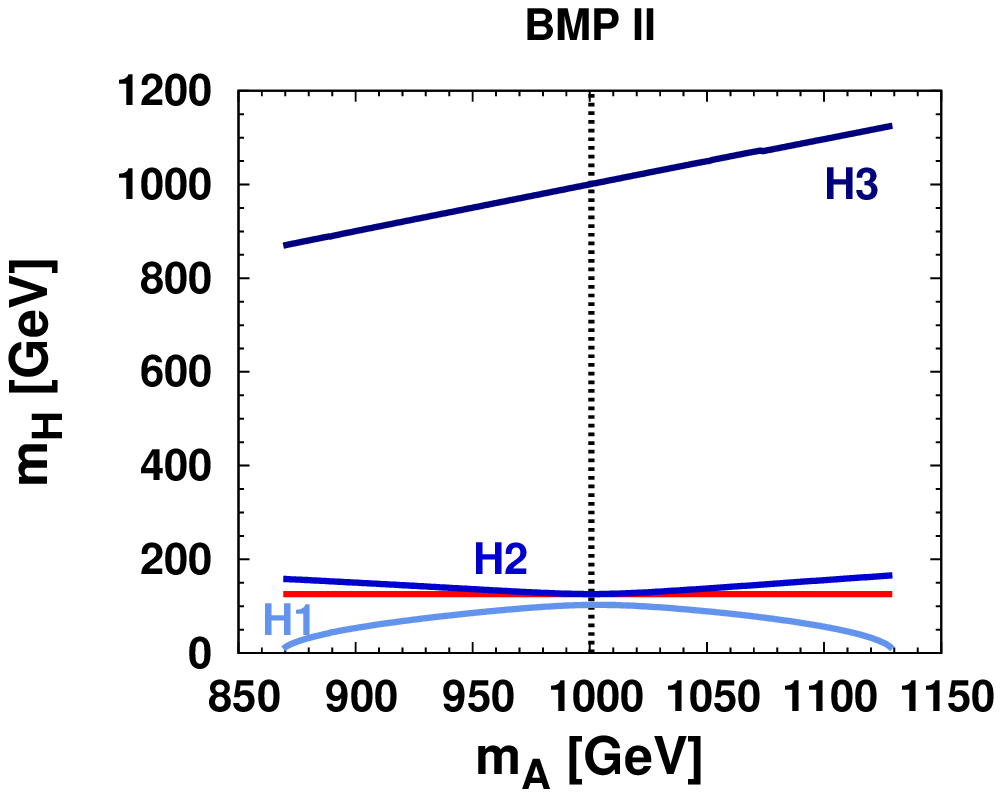} 
 \includegraphics[width=0.49\textwidth]{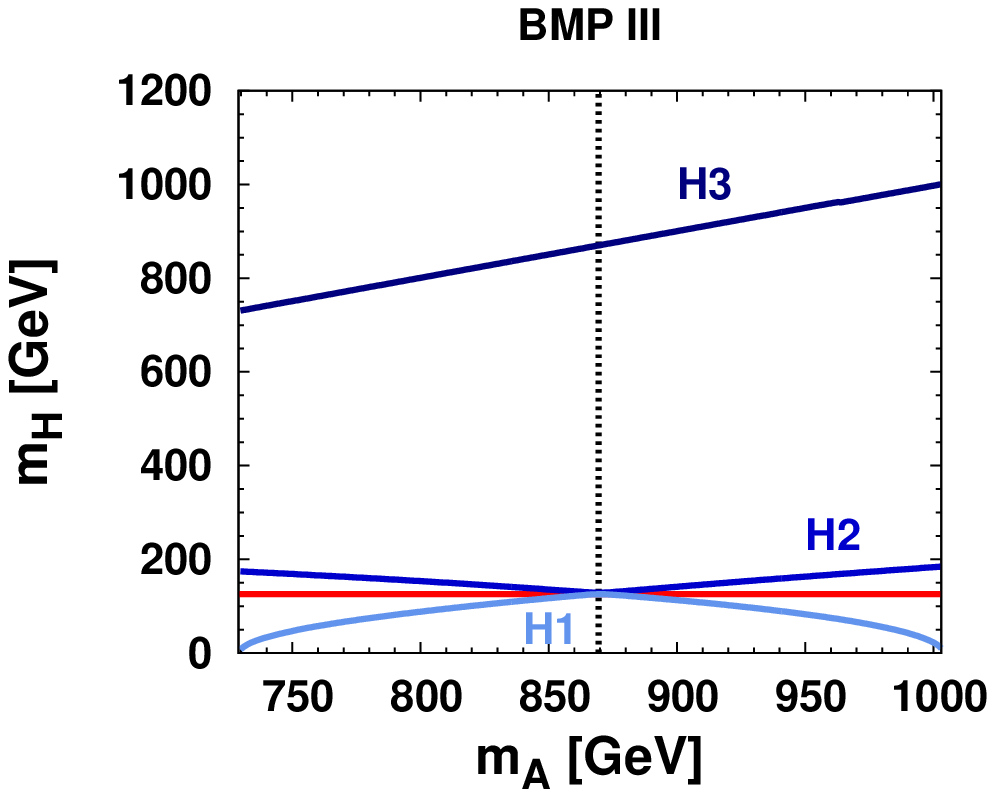} 
 \end{center}
 \caption{ 
Top left: comparison of the Higgs boson mass within the CMSSM and NMSSM plotted as function of $\mhalf$ for fixed values 
of all other parameters. 
The stop mass is indicated at the top. 
 The other panels show the three Higgs boson masses of the NMSSM Higgs mass matrix as function of $M_A$. The horizontal 
 (red) line corresponds to a Higgs boson mass of 126 GeV. The vertical line shows the value of $M_A$ which corresponds 
 to a 126 GeV Higgs with SM couplings. The corresponding NMSSM parameters are given in Table \ref{t1}. The salient 
 features of the benchmark points are: BMP I: $m_{H_2}=126$ GeV and low $H_3$ mass; BMP II: $m_{H_2}=126$ GeV and a high 
 $H_3$ mass; BMP III: the lightest Higgs boson has   $m_{H_1}=126$ GeV, which requires a large $H_3$ mass.}\label{f1}
 \end{figure}

\subsection{Benchmark scenarios with a SM-like Higgs boson }\label{mh126}

As discussed in the introduction, finding additional Higgs bosons would be the key to prove physics beyond the SM and 
the properties
of heavier Higgs bosons could distinguish between the CMSSM and NMSSM. In the CMSSM the stop mass has to be heavy enough 
to provide a mass of 
126 GeV for the lightest Higgs mass.  The decay properties of the heavier Higgs boson depend  on the masses, since new 
channels might open up. Furthermore, in the 
NMSSM one has three Higgs boson masses, so one has a choice to define which Higgs boson should have a mass of 126 GeV, 
the lightest 
one or second lightest one. In addition, the relatively large values of the $\lambda$ and $\kappa$ parameters of the 
Higgs self couplings allow for large branching ratios 
of the heaviest Higgs into the two lighter Higgses. Such couplings are absent in the CMSSM. In addition, the LSP is 
bino-like in the CMSSM, but turns out to have a large Higgsino component in the NMSSM. This leads to sizeable 
decays into gauginos (including LSPs) in the NMSSM, but these decay modes are practically absent in the CMSSM.
To study these features, we have defined four different bench mark points for similar values of  $\mzero$ and  $\mhalf$, 
chosen to yield a 126 GeV Higgs mass. The mass of the heavier Higgses is largely determined by the mass of the 
pseudoscalar Higgs boson $M_A$, both in the CMSSM and the NMSSM. However, in the CMSSM $M_A$ is 
proportional to $\mhalf$, while in the NMSSM it is independent of $\mzero$ and  $\mhalf$, as will be discussed in 
section \ref{results}, where we extend the analysis to arbitrary values of the SUSY masses. The four representative BMPs 
considered for values of  $\mzero$ and $\mhalf$ requiring a 126 GeV Higgs mass are:  the first BMP is for the CMSSM, 
while BMP I-III are for the NMSSM. In the NMSSM the mass of the heavier Higgs can be varied by the $\mu_{eff}$ parameter 
and trilinear couplings. These have been chosen such, that BMP I has a relatively light $M_A$ value, while  BMP II and 
III have heavy $M_A$ values (TeV range). The latter two distinguish themselves by the fact that for BMP II  the second 
lightest Higgs has a mass of 126 GeV, while for BMP III the lightest Higgs boson mass is 126 GeV.

The characteristics of the BMPs are displayed in Fig. \ref{f1}.  The top left panel shows the Higgs mass of the observed 
Higgs boson as function of $\mhalf$.
Increasing $\mhalf$ increases the stop mass, as indicated on the scale at the top of the fi\-gure. 
One notices the steep increase in the Higgs boson mass for the NMSSM,  which reaches 126 GeV for significantly lower 
stop masses than in the CMSSM.  The main reason is that an increase in $\mhalf$ also increases $M_A$ and in the NMSSM 
the light Higgs boson masses depend on $M_A$ via the off-diagonal elements in the mass matrix (Eq. \ref{mix}). The $M_A$ 
dependence of the three eigenvalues of the Higgs mass matrix 
is shown  in the top right panel of Fig. \ref{f1}. One observes that the second heaviest Higgs boson can have a mass 
close to 126 GeV for pseudo-scalar Higgs boson masses in the range around 300 GeV.
The allowed range of $M_A$ is limited by the requirement that all Higgs boson values have to be positive, but they can  
be shifted by different values of $\mu_{eff}$, $A_{\lambda}$, $\tb$ and ${\kappa}$ (see Eq. \ref{MA}). This is shown in 
the bottom row of Fig. \ref{f1}, where  larger values of $A_{\lambda}$ have been used.  The parameters for the different 
plots, indicated as benchmark points BMP I-III, are given in Table \ref{t1}.
 The corresponding SUSY masses are given in Table  \ref{t2} and the Higgs masses with the branching fractions in Tables 
 \ref{t3} and \ref{t4}. The branching fractions depend on the masses and mixings and will be discussed later.

\begin{table}
\footnotesize 
\centering
\begin{tabular}{l|lllllllll}
\hline\noalign{\smallskip}
  Input&$m_0$&$m_{1/2}$  &$A_0$ &$\tb$ &$A_{\kappa}$ &$A_{\lambda}$&$\kappa_{SUSY}$ &$\lambda_{SUSY}$ &$\mu$   \\
\noalign{\smallskip}\hline\noalign{\smallskip}
CMSSM 	& 2500	& 2375	& -4999	& 48.11	& -	& -	& -	& -	& 3339\\
BMP I 	& 2400 	& 550	& -976	& 2.69	& -848	& -509	& 0.38	& 0.65	& 120\\
BMP II	& 2450 	& 550 	& -1840	& 4.18	& 2549 	& 1774 	& 0.12	& 0.68	& 229\\
BMP III	& 2400 	& 600 	& -1902	& 3.08	& 1206 	& 756 	& 0.14	& 0.64	& 263\\
\noalign{\smallskip}\hline
\end{tabular}
\caption{List of GUT scale input parameters for several benchmark points in the CMSSM and NMSSM, respectively. The top 
and bottom masses are $m_t=172.5$ GeV and $m_b=4.25$ GeV;  $sgn(\mu)=+1$ for both models. The input parameters have been chosen to yield a Higgs boson mass of 126 GeV and have SUSY masses just outside the excluded regions in Fig. \ref{f4}. 
$\kappa$ and $\lambda$ are given at the SUSY scale as input. The GUT scale values for BMP I(II,III) are 
$\kappa_{GUT}=3.04(0.35,0.34)$ and $\lambda_{GUT}=2.16(1.46,1.22)$. }
\label{t1}
\end{table}

\begin{table}
\footnotesize 
\centering
\begin{tabular}{l|lllllllllll}
\hline\noalign{\smallskip}
&$\chi^0_1$&$\chi^0_2$&$\chi^0_3$&$\chi^0_4$&$\chi^0_5$&$\chi^\pm_1$&$\chi^\pm_2$&$\tilde{t}_1$&$\tilde{t}_2$&$\tilde{b}_1$&$\tilde{b}_2$\\
\noalign{\smallskip}\hline\noalign{\smallskip}
CMSSM	& 1075	& 1993	& 3332	& 3334	& -	& 1993	& 3334	& 3263	& 3908	& 3902	& 4014\\
BMP I 	& 76	& 160	& 197	& 248 	& 477	& 109	& 477	& 992	& 1925	& 1918	& 2552\\
BMP II	& 80 	& 212	& 266	& 271	& 484	& 217	& 484	& 983	& 1957	& 1947	& 2596\\
BMP III	& 119	& 232	& 293	& 297	& 527	& 249	& 526	& 960	& 1952	& 1942	& 2588\\
\noalign{\smallskip}\hline
\end{tabular}
\caption{The masses of neutralinos, charginos, stops  and sbottoms in GeV for the CMSSM and the three NMSSM  benchmark 
points (BMP I-III) in Table \ref{t1}.  
}
\label{t2}
\end{table}

\begin{table}
\footnotesize 
\centering
\begin{tabular}{l|ccc||ccccc}
\noalign{\smallskip}\hline\noalign{\smallskip}
\multicolumn{9}{c}{Branching Ratios [\%]}\\
\hline\noalign{\smallskip}
  			&\multicolumn{3}{c}{ CMSSM }& \multicolumn{5}{c}{NMSSM (BMP I)}  \\
\noalign{\smallskip}\hline\noalign{\smallskip}
			& h	& H	& A	& $H_1$	& $H_2$	& $H_3$	& $A_1$	& $A_2$\\
Mass [GeV]		& 126	& 2256	& 2256 	& 86	& 126	& 336	& 214	& 325\\
\noalign{\smallskip}\hline\noalign{\smallskip}
$b \bar{b} $ 		& 67.6	& 85.2 	& 85.2	& 90.6	& 63.6 	& 3.0	& 0.2	& 1.9	\\
$W^{+}W^{-} $  		& 17.7	& 1.7e-5& -	& 6.5e-7& 19.6	& 0.2	& - 	& -	\\
$ \tau \tau $  		& 5.0 	& 14.6	&14.6	& 8.8	& 6.5	& 0.4	& 0.02 	& 0.2	\\
$h h$  			& -	& 8.9e-5& -  	& -	& - 	& -	& - 	& -	\\
$H_1 H_2 $  		& - 	& -	& -  	& -	& -	& 41.9	& - 	& - 	\\
$A_1 H_1 $  		& -	& -	& -	& -	& - 	& -	& -	& 4.0 	\\
$ Z h $  		& -	& - 	& 1.7e-5& -	& -	& - 	& - 	& -	\\
$ Z H_1 $  		& -	& - 	& -	& -	& -	& - 	& 0.3 	& 26.8	\\
$\chi_1^0 \chi_1^0$  	& - 	& 4.7e-5& 5.3e-4& - 	& -	& 5.7	& 99.5	& 38.1	\\
$\chi_1^0 \chi_3^0$	& -	& -	& -	& -	& - 	& 20.8	& -	& 4.2	\\
$ \chi_1^{+} \chi_1^{-}$& - 	& -	& - 	& -	& -	& 20.7	& -	& 18.4 	\\
\noalign{\smallskip}\hline\noalign{\smallskip}
$\sigma_{prod}$ [pb] 	& 19.3& 1.3e-5& 1.3e-5&2.57& 19.1& 0.57& 1.6e-2& 0.41\\
\noalign{\smallskip}\hline
\end{tabular}
\caption{Neutral Higgs boson masses and branching ratios in the CMSSM and NMSSM (BMP I). Note the different branching 
fractions in the NMSSM for the heaviest scalar $H_3$ and pseudo-scalar Higgs boson $A_2$ compared to the CMSSM ($H$ and 
$A$): in the CMSSM the decay into b-quarks is enhanced, because of the large values of tan$\beta$ (see Fig. \ref{f3}). 
In addition the lightest neutralino has a large singlino  component, which leads to large branching ratios of $A_2$ into 
$Z+H_1$ and $H_1+H_1$ (via the $\lambda$ couplings in Eq. \ref{NMSSM}) and appreciable decays into neutralinos for this 
BMP with small $\mu_{eff}$, which implies low masses for  the LSPs ($A_1$ decays  nearly 100\% into LSPs).  This leads 
to  signatures with large missing energy for   Higgs decays in the NMSSM. The cross section represents the Higgs 
production cross section at 8 TeV for the dominant gluon-gluon fusion process. }
\label{t3}
\end{table}

\begin{table}
\footnotesize 
\centering
\begin{tabular}{l|ccccc||ccccc}
\noalign{\smallskip}\hline\noalign{\smallskip}
\multicolumn{11}{c}{Branching Ratios [\%]}\\
\hline\noalign{\smallskip}
  &\multicolumn{4}{c}{ NMSSM    (BMP II) }&& \multicolumn{5}{c}{NMSSM (BMP III)}  \\
\noalign{\smallskip}\hline\noalign{\smallskip}
	&$H_1$&$H_2$&$H_3$&$A_1$&$A_2$&$H_1$&$H_2$&$H_3$&$A_1$&$A_2$\\
Mass   [GeV] & 103 &126&1001&91&1001&126&129 & 870&118&869\\
\noalign{\smallskip}\hline\noalign{\smallskip}
$ b \bar{b} $ 		& 90.5 	& 61.9 	& 0.9 	& 90.9 	& 0.9 	& 61.7 	& 88.6 	& 0.7 	& 90.4 	& 0.6 	\\
$ t \bar{t} $ 		& 0.0  	& 0.0  	& 9.6 	& 0.0  	& 10.4	& 0.0  	& 0.0  	& 22.1	& 0.0  	& 23.5	\\ 
$ \tau \tau $ 		& 9.1  	& 6.4  	& 0.1 	& 8.8  	& 0.1 	& 6.3  	& 9.3  	& 0.1 	& 9.3  	& 0.1 	\\
$W^{+}W^{-} $ 		& 1.2e-4& 20.6 	& 1.7e-4& - 	& - 	& 20.6 	& 1.7  	& 7.9e-3& - 	& -   	\\
$ \chi_1^0 \chi_1^0$    & - 	& -   	& 10.7 	& -   	& 11.8 	& -   	& -   	& 9.0  	& -  	& 9.6 	\\
$\chi_1^0 \chi_3^0$    	& - 	& -   	& 5.1 	& -   	& 6.3 	& -   	& -   	& 5.5  	& -  	& 8.9 	\\ 
$\chi_1^{+} \chi_1^{-}$	& - 	& -  	& 3.2 	& -  	& 5.9	& -  	& -  	& 2.4 	& -	& 6.3 	\\
$H_1 H_2 $  		& - 	& - 	& 14.8 	& - 	& -  	& -  	& -  	& 13.6 	& - 	& -  	\\
$A_1 H_2 $  		& - 	& - 	& -     & - 	& 13.5 	& -  	& -  	& -  	& - 	& 0.2 	\\
$Z A_1 $    		& - 	& - 	& 12.3  & - 	& - 	& -  	& -  	& 10.6 	& - 	& - 	\\
$Z H_1 $   		& - 	& - 	& -     & - 	& 13.6	& - 	& -  	& -   	& - 	& 0.04 	\\
$A_1 H_1 $  		& - 	& - 	& -     & - 	& 0.3	& -	& -	& -	& -	& 11.7	\\
$Z H_2 $    		& - 	& - 	& -     & - 	& 8.1e-4& -	& -	& -	& -	& 11.9	\\
\noalign{\smallskip}\hline\noalign{\smallskip}
$\sigma_{prod}$ [pb] 	& 0.33	& 19.3	& 1.6e-3& 0.13	& 1.9e-3& 19.5	& 3.9e-2& 7.1e-3& 1.7e-2&7.6e-3	\\
\noalign{\smallskip}\hline
\end{tabular}
\caption{ Neutral Higgs boson masses and branching ratios in the NMSSM for BMP II and III.  A heavy Higgs boson of about 
1 TeV can be reached by choosing either large values of $\mu_{eff}$ or large values of tan$\beta$ combined with large 
values of $A_{\lambda}$. This leads to different allowed regions of $A_{\kappa}$. If the mass of the lightest 
pseudo-scalar Higgs boson is high enough, it decays mostly into neutralinos, as shown before in Table \ref{t3}. This is not possible for a light $A_1$, 
which decays into b-quarks. The heavier masses for $H_3$ and $A_2$ change the branching fractions compared to Table 
\ref{t3}, because $t\overline{t}$  becomes kinematically allowed. As in the previous table, the cross section 
corresponds to the Higgs production cross section at 8 TeV for the dominant gluon-gluon fusion process.}
\label{t4}
\end{table}

\begin{table}
\footnotesize 
\centering
\begin{tabular}{l|ccccc||ccc}
\hline\noalign{\smallskip}
&&\multicolumn{3}{c}{Higgs mixing [\%]}& &\multicolumn{3}{c}{Reduced couplings}\\
\noalign{\smallskip}\hline\noalign{\smallskip}
& & $H_d$ & $H_u$ & S && $\kappa_{u}$ & $\kappa_{d}$ & $\kappa_{W,Z}$\\
\noalign{\smallskip}\hline\noalign{\smallskip}
		&\multicolumn{1}{c|}{$H_1$} &26.9&-11.4&95.6&&-0.12&0.77&-0.01\\
		&\multicolumn{1}{c|}{$H_2$} &36.5&93.1&0.8&&0.99&1.04&1.00\\
	BMP I 	&\multicolumn{1}{c|}{$H_3$} &89.1&-34.7&-29.3&&-0.37&2.55&-0.01\\
	$\sim $ BMP II	&\multicolumn{1}{c|}{$A_1$} & -14.6&-5.4 & 98.8&& -0.06& -0.42& 0.00\\
		&\multicolumn{1}{c|}{$A_2$} & 92.6& 34.4&15.6 && 0.37& 2.66& 0.00\\
\noalign{\smallskip}\hline\noalign{\smallskip}
\noalign{\smallskip}\hline\noalign{\smallskip}
		&\multicolumn{1}{c|}{$H_1$} & 30.1& 95.0& -8.2&& 0.99&0.98 &0.99 \\
		&\multicolumn{1}{c|}{$H_2$} & 12.7& 4.6& 99.1 && 0.05& 0.41&0.08\\
	BMP III	&\multicolumn{1}{c|}{$H_3$} & 94.5& -30.9& -10.7&& -0.32& 3.18& 0.003\\
		&\multicolumn{1}{c|}{$A_1$} &  -9.3& -3.0 & 99.5 &&  -0.03& -0.30&0.00 \\
		&\multicolumn{1}{c|}{$A_2$} &  94.7&  30.7&  9.8&& 0.32& 3.07&0.00\\
\noalign{\smallskip}\hline
\end{tabular}
\caption{The NMSSM scalar and pseudo-scaler Higgs boson mixings   and the reduced couplings, i.e. the couplings in units 
of the SM couplings for the benchmark points  given in Table \ref{t1}. The couplings and mixings of BMP I and II are 
similar, so only  the couplings  for BMP I are shown. One observes that one of the lightest Higgs boson has couplings 
close to the SM Higgs boson, while the other light Higgs boson hardly couples to the gauge bosons, thus explaining why 
it could not have been discovered at LEP, but it  couples preferentially to d-type fermions, so it will decay into 
b-quarks and $\tau$-leptons.   Also the heaviest Higgs boson has small couplings to the gauge bosons, as in the CMSSM. 
One observes that $H_1$ and $H_2$ have quite different reduced couplings, so  deviations from SM couplings for the 126 
GeV boson are readily obtained by changing the mixing, i.e. moving further away from the maxima in the $H_1$ curves in 
Fig. \ref{f1}. }
\label{t5}
\end{table}

In the NMSSM either the lightest Higgs boson (by definition $m_{H_1}$) or the second lightest $m_{H_2}$ can be 126 GeV 
with SM-like couplings. These couplings were required to be within 10\% equal to the SM couplings. Possible deviations 
from the SM couplings can be obtained by different mixings in the Higgs sector, as will be discussed below. To obtain a 
significant contribution to the Higgs boson mass at tree level from the mixing - thus preventing multi-TeV stop masses - 
the masses $m_{H_1}$ and $m_{H_2}$ have to be rather close, so one expects in the NMSSM a second Higgs boson either 
below or above 126 GeV, as is apparent from Fig. \ref{f1} for the different benchmark points BMP I-III.  In the maximum 
(minimum) of $H_1$ ($H_2$) the mixing is minimal and away from these extremes the mixing increases, thus lowering one 
Higgs mass and increasing the other. In BMP I and II $H_2$ has SM-like couplings, while in BMP III $H_1$ has SM-like 
couplings, i.e. the reduced couplings, defined as the ratio of the Higgs couplings to SM couplings, are close to 1, so 
the SM-like Higgs can either decrease or increase by the mixing, depending on which side one is from the maximum 
(minimum) of $H_1$ ($H_2$) in Fig. \ref{f1}  (shift-down or shift-up, respectively).

 The Higgs mixings and the  reduced couplings for the BMPs are shown in Table \ref{t5}. The corresponding SUSY 
 parameters and Higgs masses  have been shown before in Tables \ref{t1} and \ref{t3},\ref{t4} , respectively.  One 
 observes that $H_1$ and $A_1$ are almost purely singlets in BMP I and II, while  $H_2$  has indeed couplings close to 
 the SM Higgs boson. $H_1$ and  $H_3$ couple preferentially to down-type fermions and have negligible couplings to 
 $W^{\pm}$- and $Z^0$-bosons. This negligible coupling of $H_1$ to the $Z^0$-boson explains why it could not have been 
 discovered at LEP, even if its mass is below the LEP limit of 114 GeV. In BMP III the roles of $H_1$ and $H_2$ are 
 interchanged, but their masses are close (126 and 129 GeV, respectively, see Table \ref{t4}).
 Note that $H_1$ and $H_2$ have quite different reduced couplings, so  deviations from SM couplings, like a different 
 value for $BR(H_2\rightarrow \gamma\gamma)$, are readily obtained by changing the mixing, i.e. moving further away from 
 the maxima in the $H_1$ curves in Fig. \ref{f1}. 
 
 The three benchmark points  yield similar $\chi^2$ values for the fits discussed below, but have different mass 
 predictions for the remaining Higgs bosons besides the 126 GeV Higgs boson, as was shown in Tables \ref{t3} and 
 \ref{t4}  before. Especially, in the CMSSM the heavier Higgs masses are all around 2.2 TeV, which is a consequence of 
 the heavy stop mass  required to be well above 3 TeV for a SM-like Higgs boson of 126 GeV. In the NMSSM no heavy stop 
 masses are required, so the heavy Higgs masses can be as low as about 280 GeV, but large values  are allowed as well, 
 as shown by the difference between BMP I and II or III  in Fig. \ref{f1}.  However, heavy Higgs masses can only be 
 obtained by a careful finetuning of large  values of $A_{\kappa}$, $A_{\lambda}$ and $A_0$, which allows to increase 
 the values of $\mu_{eff}$ and thus the mass of $H_3$, as can be seen from  Eq. \ref{MA}. It is especially important to 
 have opposite signs for $A_0$  and $A_\kappa, A_\lambda$ in order to get large values of $m_{H_3}$. The value of 
 $\mu_{eff}$  is determined by the vev of the singlet, so it is independent of $m_0$ and $m_{1/2}$ in contrast to the 
 CMSSM, where the $\mu$ parameter is effectively proportional to  $\mhalf$ after EWSB is imposed.
The heavy $H_3$ scenarios are  possible, but will require high luminosity, because of the small cross sections 
(proportional to $1/m_{H_3}^{4}$), as can be seen from the production cross section in Table \ref{t3} and \ref{t4}.

One observes that the heavier Higgs boson $H$ in the CMSSM decays preferentially into b-quarks and $\tau$-leptons, while 
the heavier Higgs boson in the NMSSM $H_3$ decays preferentially into $H_2$ + $H_1$ and gauginos.  The Higgs boson with 
a large singlet component is hard to discover at the LHC because of its reduced couplings, but it may be discovered in 
the decay mode of  the heavier Higgs boson $H_3$. Given the Higgs self-coupling one can expect a large fraction of the 
decay of the heavier Higgs into the two lighter Higgses \cite{Kang:2013rj}. Note that this would lead to double Higgs 
boson production, i.e. events with two Higgs bosons, where  one of them has a mass of 126 GeV. This would be a unique 
NMSSM signature. Such large branching ratios are indeed expected for BMP I, as indicated in Table \ref{t3}. In contrast, 
double Higgs production in the CMSSM is negligible, as shown on the left side of the table. Another interesting 
signature is given by the lightest pseudo-scalar Higgs in BMP I. Since the LSP in the NMSSM is higgsino-like, the 
lightest pseudo-scalar Higgs boson  $A_1$ decays almost exclusively to two LSPs,  i.e. invisible decays, at least if 
kinematically allowed, which is not the case for BMP II and III with  lighter $A_1$ bosons.
These benchmark points mass have much larger values of $A_{\kappa}$ (Table \ref{t1}) and a correspondingly lighter  
$A_1$ boson. In this case the decays into two b-quarks and $\tau$-leptons prevail, as indicated in Table \ref{t4}. 
 In case of heavy $H_3$ scenarios the $t\overline{t}$ decay modes are allowed and the decays of heavier into lighter 
 Higgs bosons or LSPs decreases, as shown in Table \ref{t4}.

\section{Experimental Constraints and Fitting Method}\label{multi}
 
\begin{figure} 
 \begin{center}
 \includegraphics[width=0.49\textwidth]{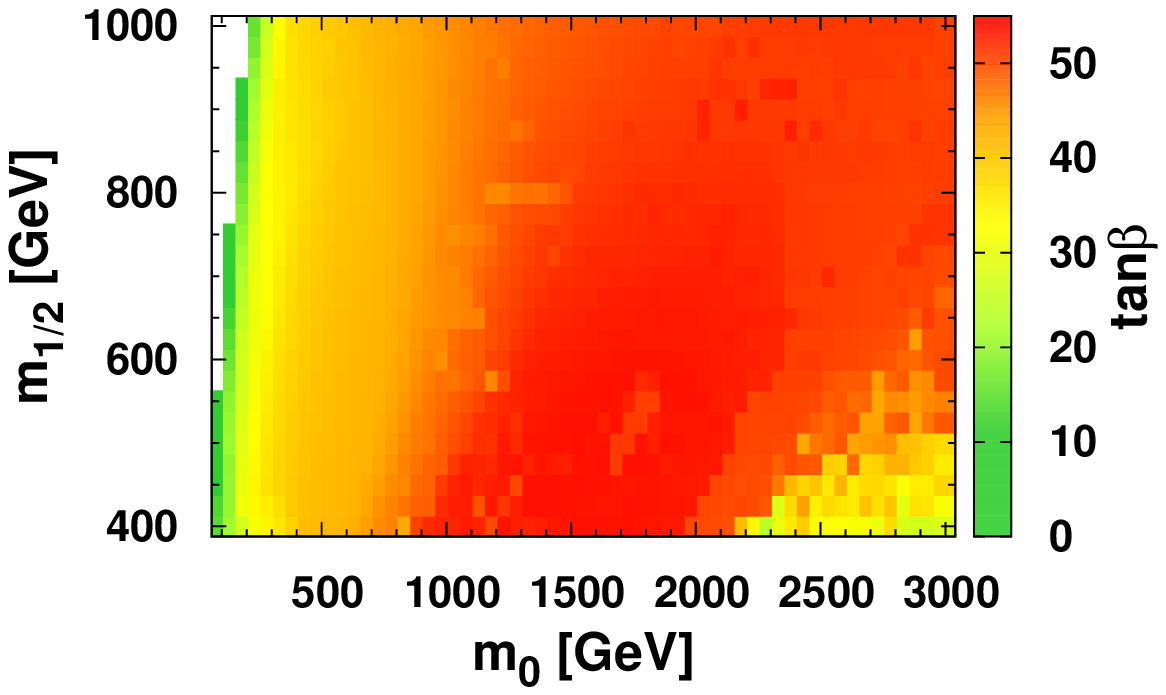}
 \includegraphics[width=0.49\textwidth]{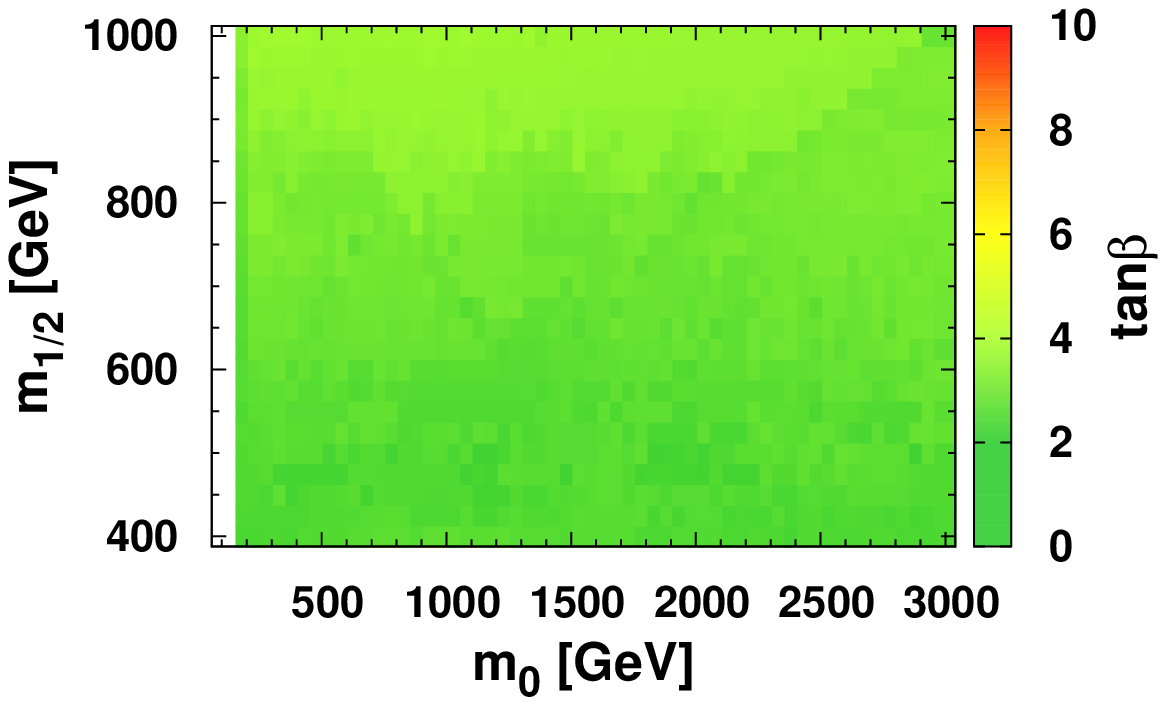} 
 \end{center}
 \caption{Optimized values of $\tb$ within the CMSSM (left) and the NMSSM (right). One can see that high values of $\tb$ 
 are preferred within the CMSSM except for the edges (co-annihilation regions), as discussed in Paper I. Low values are 
 favored within the NMSSM.}\label{f3}
 \end{figure}

Instead of concentrating on a few benchmark points we combine the data on the observed Higgs boson with a mass of 126 
GeV with all cosmological and other accelerator constraints by minimizing the $\chi^2$ function for each pair of 
$\mzero,\mhalf$ in a lattice of these variables,
so for each set of SUSY masses only the couplings have to be optimized.  The couplings for each lattice point can be 
fitted independently, so the problem can be parallelized and  solved fast on a cluster of parallel processors. Details 
and values of the experimental constraints have been discussed previously in Ref. \cite{Beskidt:2012sk}, which we call 
Paper I. Here only the CMSSM was considered. Concerning the experimental constraints we  have replaced the upper limit 
on the branching ratio $\bsmm$  with the recently observed values  given in Ref. \cite{LHCbbsmm,CMSbsmm}, which have a 
significance of more than 4$\sigma$ for each experiment and are consistent with the SM values. Since the SUSY 
contributions to this branching ratio are proportional to $\tan^6\beta$, it starts to exclude regions of the CMSSM 
parameter space, since here large values of $\tan\beta$ are required, but it has no influence for the NMSSM, since here 
small values of $\tan\beta$ are needed (see Fig. \ref{f3} afterwards).  An updated value for the  relic density from the 
recent Planck data has been used \cite{Ade:2013lta}. As in Paper I we use the multi-step fitting technique, which means 
that one first fits the most important couplings, in this case $\lambda$, $\kappa$ and $\tb$. All other parameters are 
optimized in a second step and the whole procedure is used iteratively, i.e. the best output values of a fit are used as 
input for the next iteration. This multi-step fitting technique usually finds  larger allowed regions than Markov Chains 
or other scanning techniques, since in case of strongly correlated parameters it is hard to find the allowed regions, 
unless one uses a correlation matrix during the scanning, which tells if parameter $i$ moves to a certain value, all 
other correlated parameters should move to specific values as well. Since this matrix is not know explicitly, it is not 
used in Markov Chains. However, in  the multi-step fitting technique all correlations are calculated during the 
minimization by using Minuit as minimizer \cite{James:1975dr}.

All observables  were calculated with the public code NMSSMTools \cite{Das:2011dg}. It has the option to define the mass 
parameters either at the GUT scale or the electroweak scale. 
It  has an interface to micrOmegas\cite{Belanger:2010pz,Pukhov:2010px} to calculate the relic density. Observables 
sensitive to SUSY contributions, like g-2, $\bsmm$, and $\bsg$ are calculated in  NMSSMTools as well and Higgs boson 
properties are calculated inside NMSSMTools with an extended version of HDECAY\cite{Djouadi:1997yw} to include the NMSSM 
specific radiative corrections from Ref. \cite{Degrassi:2009yq}.

\begin{figure} 
 \begin{center}
 \includegraphics[width=0.49\textwidth]{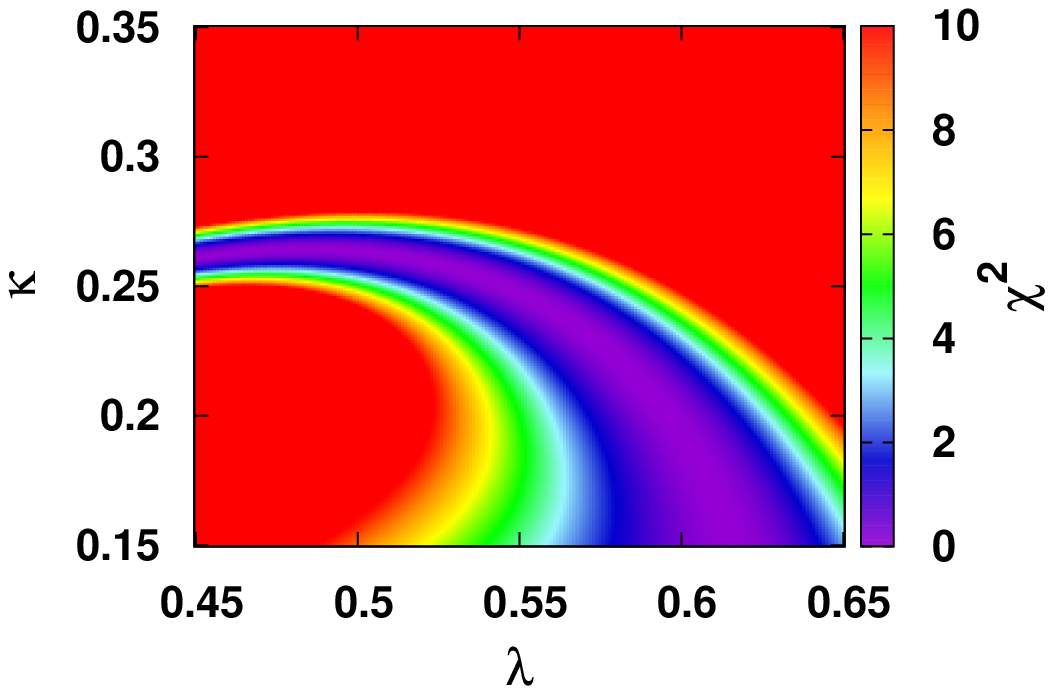}
 \includegraphics[width=0.49\textwidth]{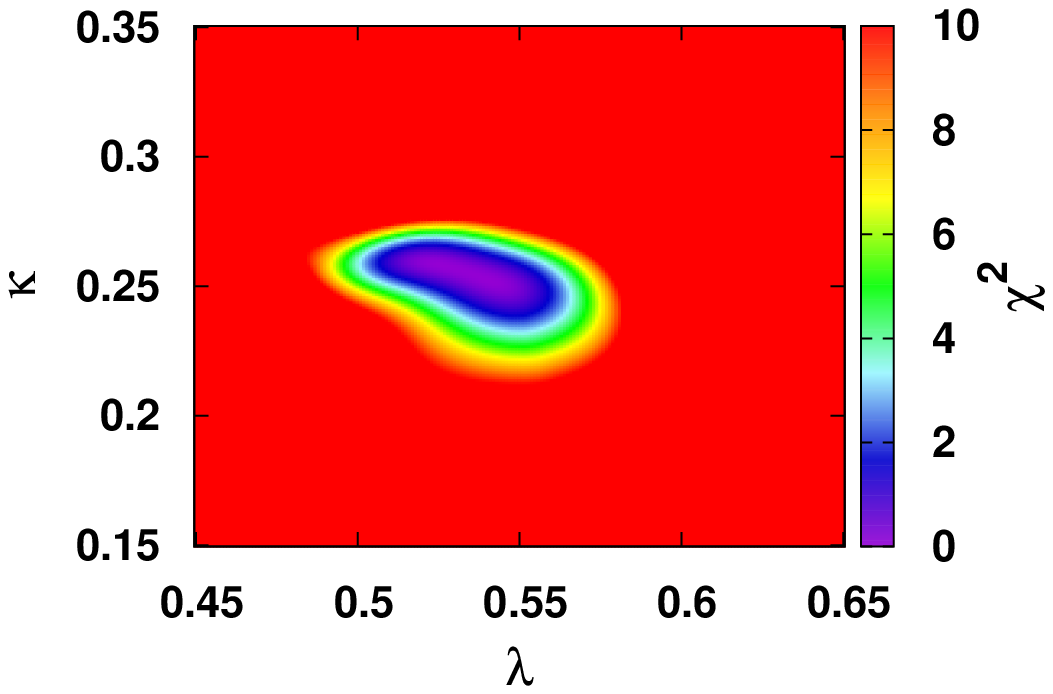} 
 \includegraphics[width=0.49\textwidth]{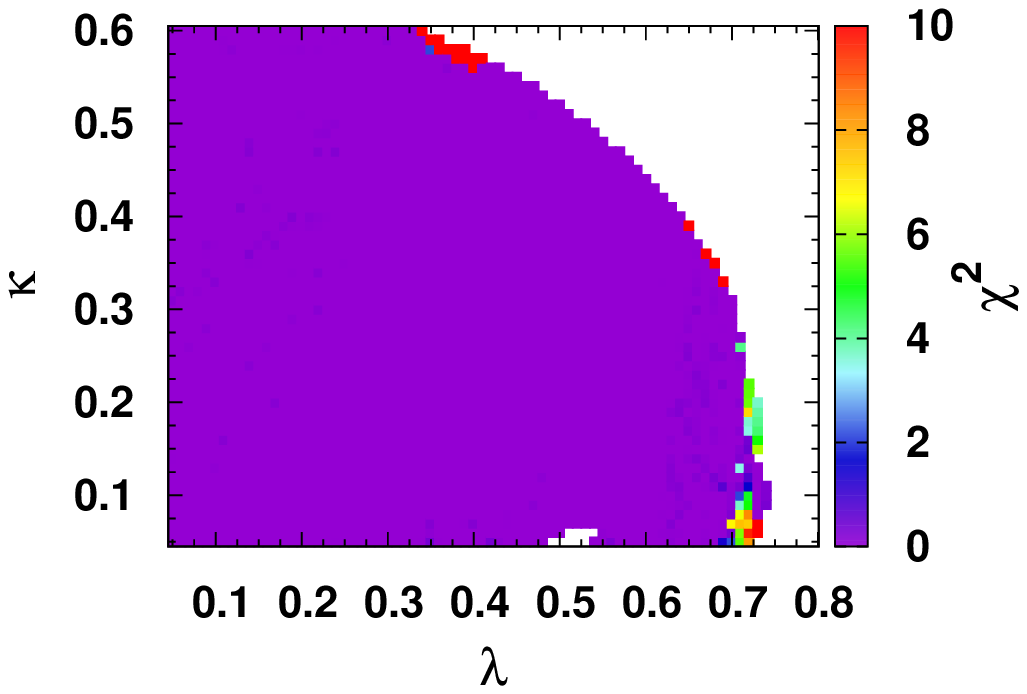}
 \includegraphics[width=0.49\textwidth]{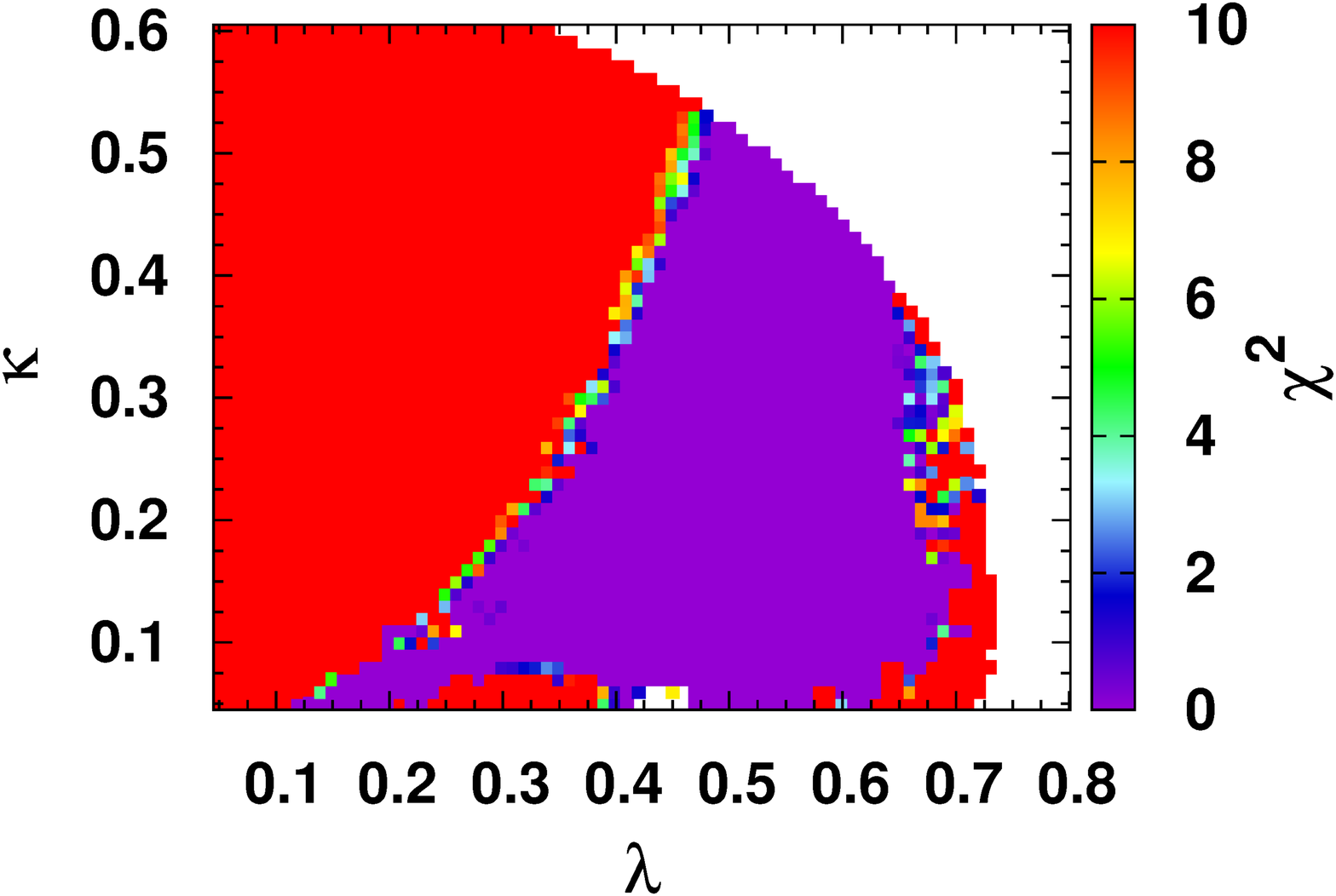} 
 \end{center}
 \caption{Top row: the $\chi^2$ distribution in the $\lambda-\kappa$ plane including the Higgs boson mass constraint 
 (left) and in addition the relic density constraint (right) for fixed other NMSSM parameters. The Higgs boson mass 
 itself does not constrain the values of $\lambda$ and $\kappa$. Only after combining the Higgs boson mass with the 
 relic density constraint leads to well defined values of the couplings (top right). Bottom row: as top row, but 
 allowing all trilinear couplings, $\mu_{eff}$ and $\tb$ to be free, which increases the allowed region to a large 
 extent  (note different scales!). }\label{f2}
 \end{figure} 

\section{Fit results}\label{results}
As mentioned above the couplings are fitted for a lattice in the $\mzero, \mhalf$ plane and the most sensitive
  couplings $\lambda$, $\kappa$  and $\tb$ are fitted first. These couplings are largely determined by the Higgs boson 
  mass and relic density, if one leaves all other parameters fixed, as shown in Fig.  \ref{f2}, top row. However, if all 
  other parameters are left free in a next step, much larger regions are allowed, as shown in the bottom row. This 
  allowed region changes again, if all constraints are included.

 Such fits can be repeated for each value of $\mzero$ and $\mhalf$ and from the  fitted couplings   all 
 other observables can be calculated in the $\mzero$,$\mhalf$ plane.  They will be compared for the CMSSM and NMSSM. 
We start with $\tb$ in Fig. \ref{f3}. One observes that in the CMSSM large values of $\tb$ are preferred in most of the 
plane.  This originates from the relic density constraint, as discussed in Paper I. In contrast, within the NMSSM low 
values of $\tb$ are preferred, as  shown in the right panel of Fig. \ref{f3}. In the NMSSM the LSP is largely 
higgsino-like and the correct relic density, or equivalently the annihilation cross section, can be obtained by 
adjusting the Higgsino content via the mixing with the singlino, so one has more freedom than in the CMSSM and 
the constraints from the relic density and direct dark matter searches can be easily met.

 \begin{figure} 
 \begin{center}
 \includegraphics[width=0.49\textwidth]{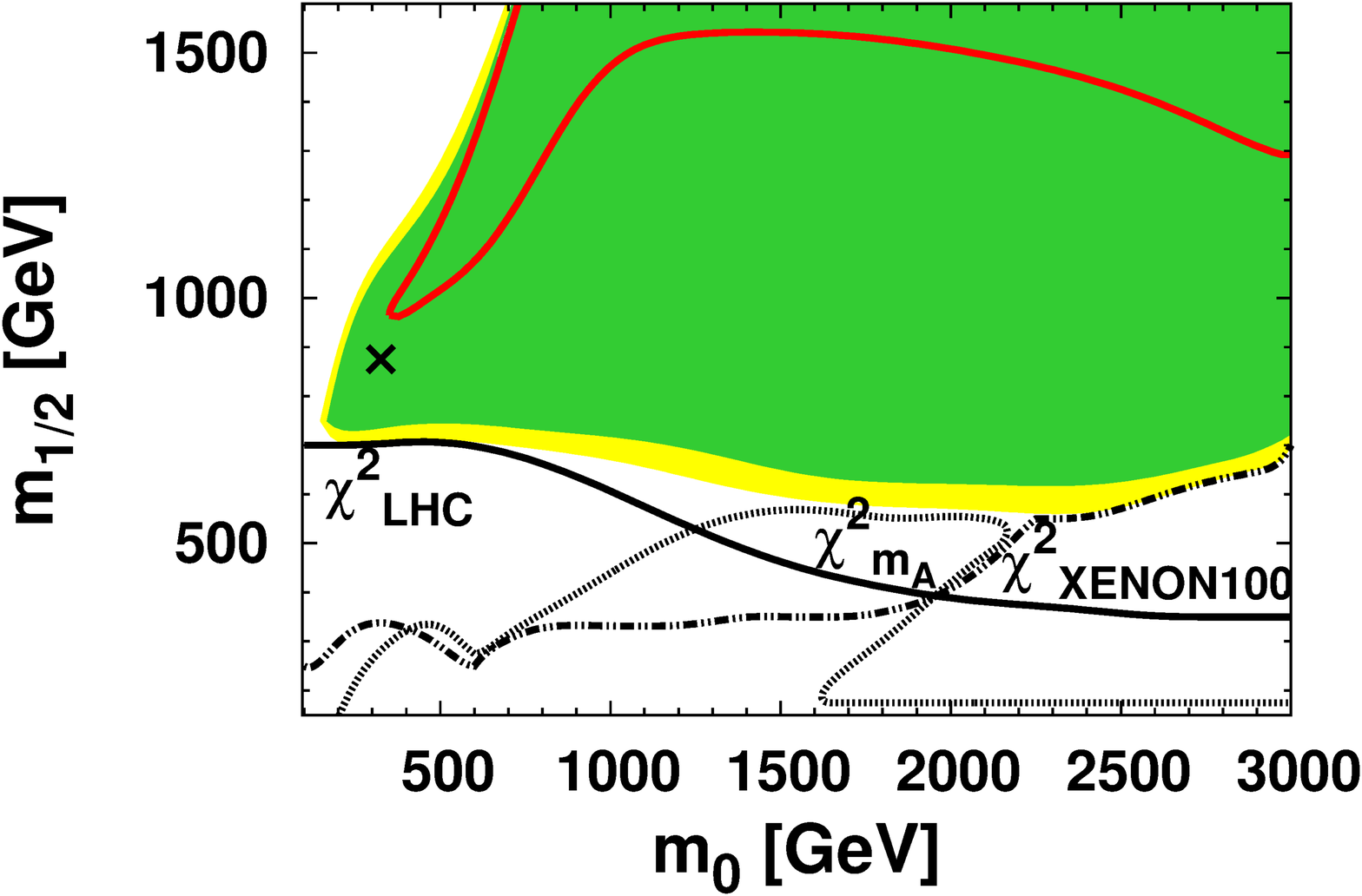}
 \includegraphics[width=0.49\textwidth]{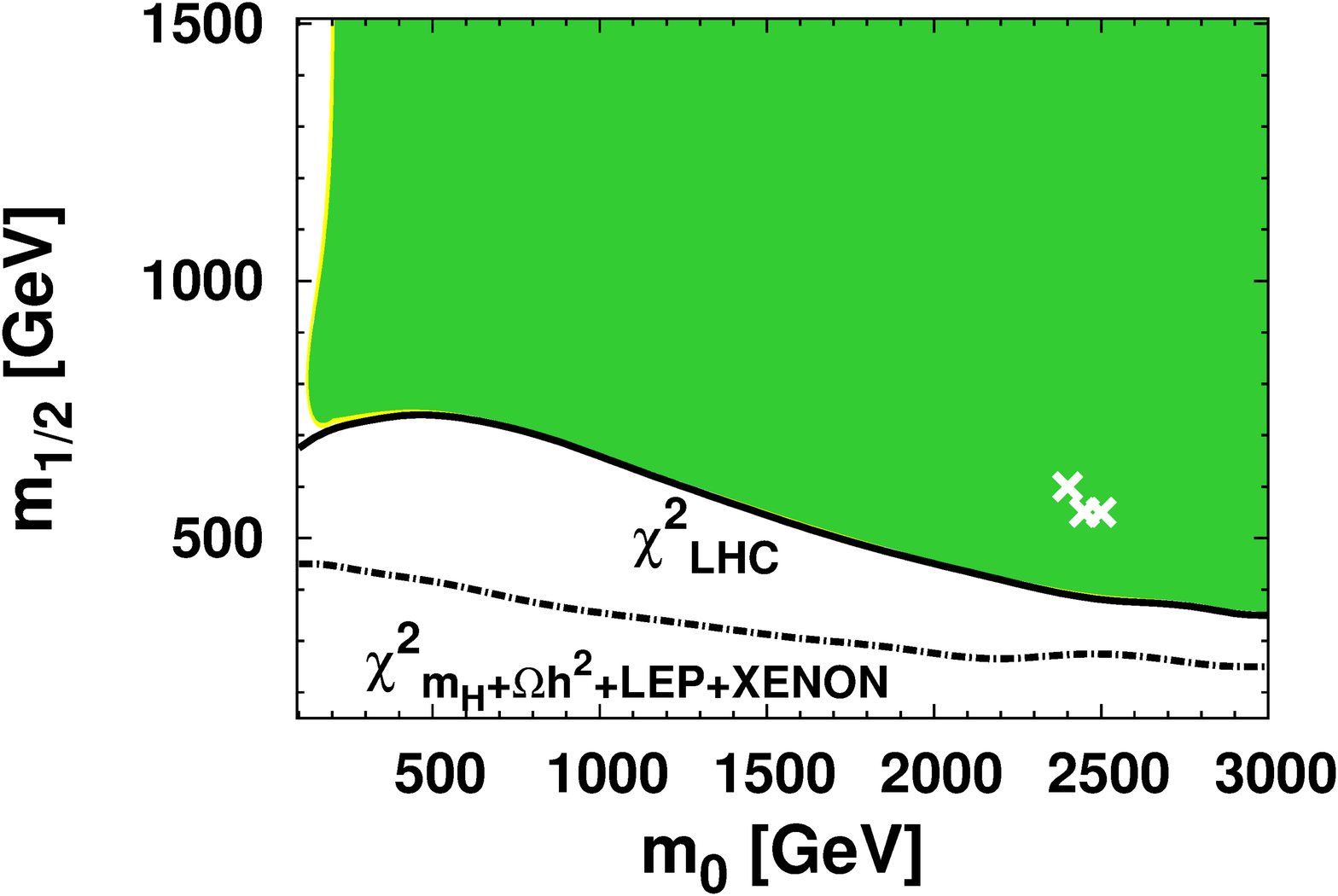} 
 \end{center}
 \caption{Excluded SUSY mass parameters, as indicated by the white region,  within the CMSSM (left) and NMSSM (right). 
 The color coding corresponds to the 1 and 2 $\sigma$ allowed regions, which means that the $\Delta\chi^2$ value is less 
 then 2.3 (green) and 5.99 (yellow), respectively. The allowed region in the left panel represents the optimization 
 including the LEP limit for the Higgs boson mass. If one includes, instead of the LEP limit the Higgs boson discovery, 
 the excluded region is dominated by the required heavy stop mass.  For  an estimated error of about 2 GeV for the Higgs 
 boson mass,  the parameter region below the solid red line in the left panel is excluded at the 95\% C.L.. The white 
 crosses in the right plot correspond to the three BMPs I-III. }\label{f4}
 \end{figure} 

\begin{figure} 
 \begin{center}
 \includegraphics[width=0.49\textwidth]{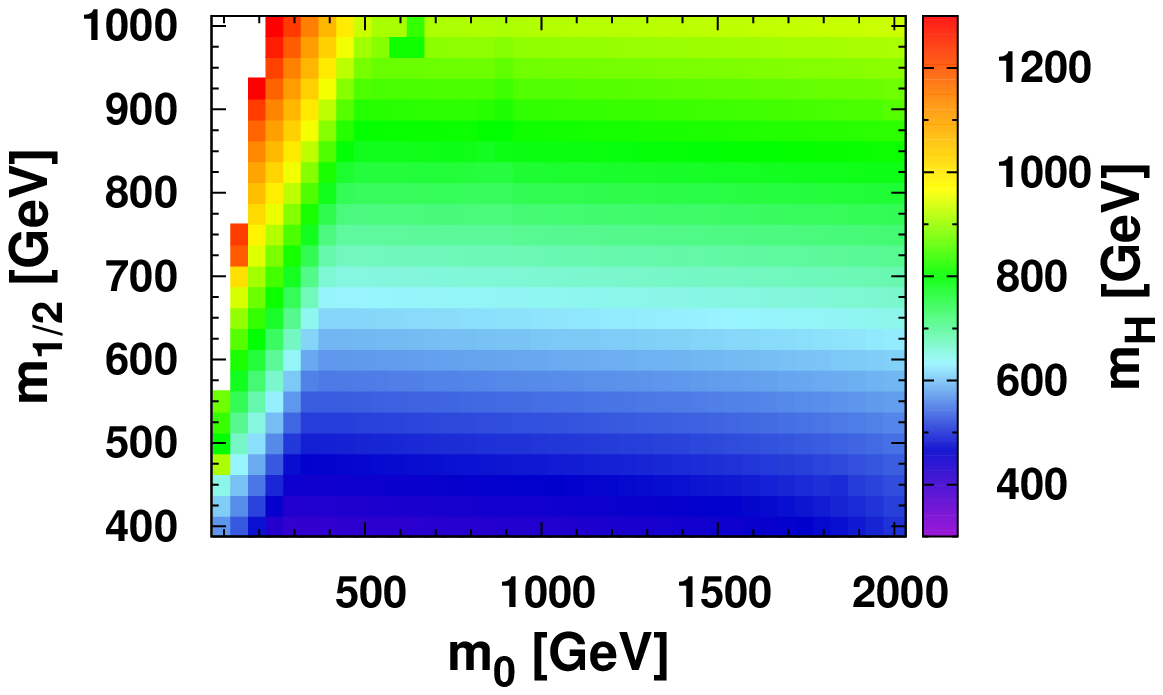}
 \includegraphics[width=0.49\textwidth]{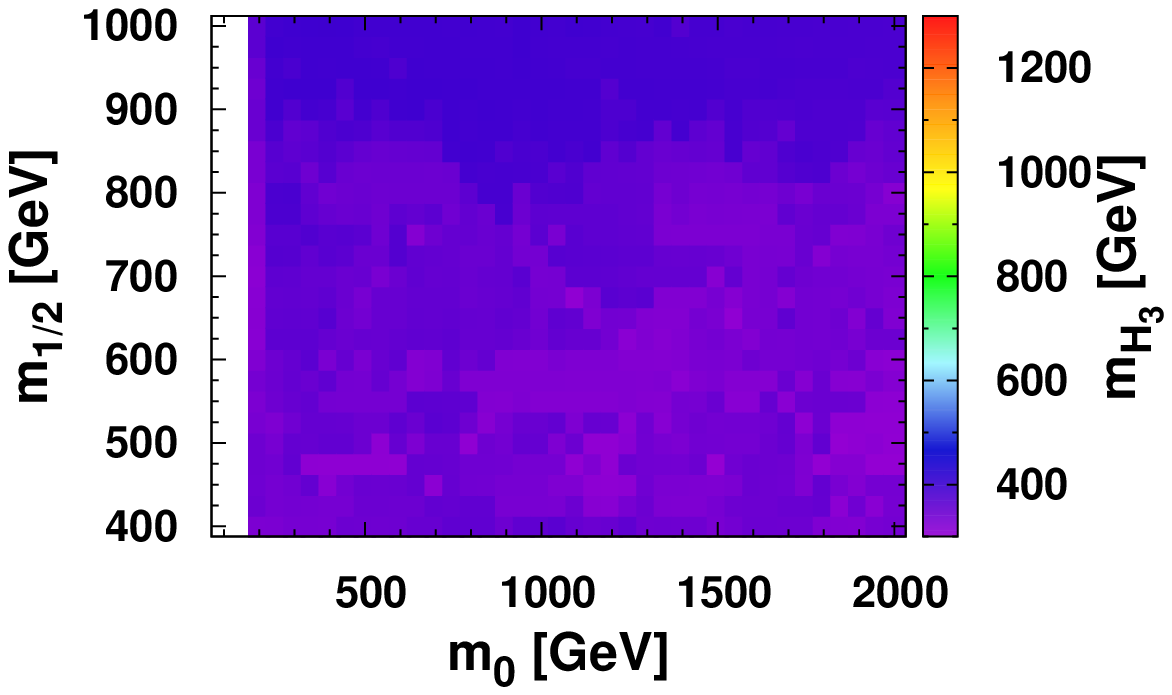} 
 \end{center}
 \caption{ 
The mass of the heavier Higgs boson in the $\mzero, \mhalf$ plane for the CMSSM (left) and NMSSM (right). The increase 
of $H$ proportional to $\mhalf$ is required by the relic density for a bino-like LSP (see Paper I), while in the NMSSM 
the LSP is higgsino-like and the Higgs masses are determined by Higgs parameters, which are largely independent of the 
SUSY masses.}\label{f5}
 \end{figure}

The SUSY masses excluded by the LHC and cosmological limits are shown for the CMSSM and NMSSM in Fig. \ref{f4}. In both 
cases the excluded region (white) is dominated by the limits from the direct SUSY searches at the LHC. For the CMSSM the 
large $\tb$ value of Fig. \ref{f3} yields in addition  contributions from the XENON100 limits, pseudo-scalar Higgs boson 
limits and $\bsmm$ observations. The latter two observables are proportional to $\tan^2\beta$ and $\tan^6\beta$, 
respectively. Details for the origin of these exclusions have been given in Paper I. As mentioned in the introduction, a 
126 GeV Higgs boson mass can only be reached for heavy stop masses, especially if one allows for the trilinear couplings  
the restricted range of the fixed point solutions from the RGEs, which do not allow maximal mixing in the stop sector. 
The excluded region depends than strongly on the assumed error in the Higgs boson mass, which is dominated by the 
theoretical uncertainty. For an error of 2 GeV the region below the red line in the left panel of Fig. \ref{f4} is 
excluded, but the best fit point ($\mzero$=1.9 TeV, $\mhalf=3.0$ TeV)   is far outside the range of the fi\-gure and 
corresponds to a stop mass above 4 TeV. For the low $\tb$ values of the NMSSM and the higgsino-like dark matter 
candidate only the LHC limits contribute, as shown in the right panel of Fig. \ref{f4}.   

For the optimized parameters in each point of the $\mzero$-$\mhalf$ plane one can calculate the Higgs boson masses and 
their branching ratios. The heaviest scalar and pseudo-scalar Higgs bosons are nearly mass degenerate in both models, 
but in the CMSSM these masses increase with $\mhalf$ to fulfill the relic density constraint (see Paper I), but in the 
NMSSM  the LSP is largely a Higgsino and through its mixing with the singlino the relic density and nucleon scattering 
cross section the experimental constraints can be easily fulfilled, as discussed above. The Higgs masses are   
determined by the additional Higgs parameters of the NMSSM, which are largely independent of $\mzero$ and $\mhalf$, so 
the Higgs masses for the benchmark points in Tables \ref{t3} and \ref{t4}   can be obtained for all points in the 
$\mzero,\mhalf$ plane. The different dependence of the Higgs masses on the SUSY masses for the CMSSM with a bino-like 
LSP and the NMSSM with a higgsino-like LSP is demonstrated in Fig. \ref{f5}.

\begin{figure} 
 \begin{center}
 \includegraphics[width=0.49\textwidth]{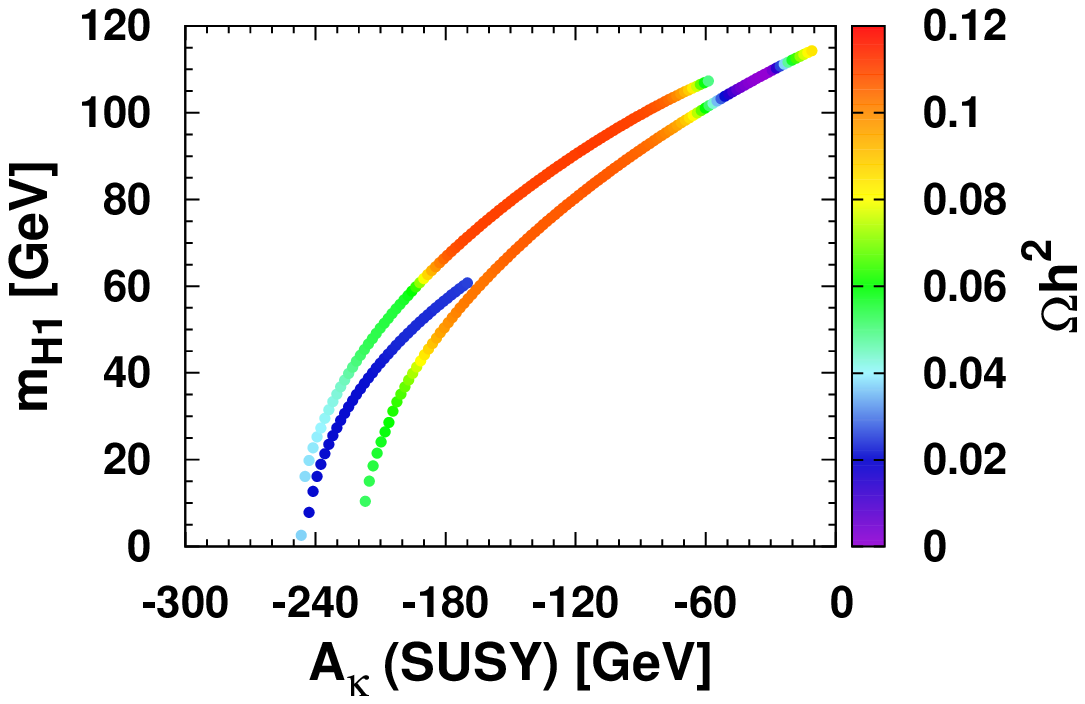}
 \includegraphics[width=0.49\textwidth]{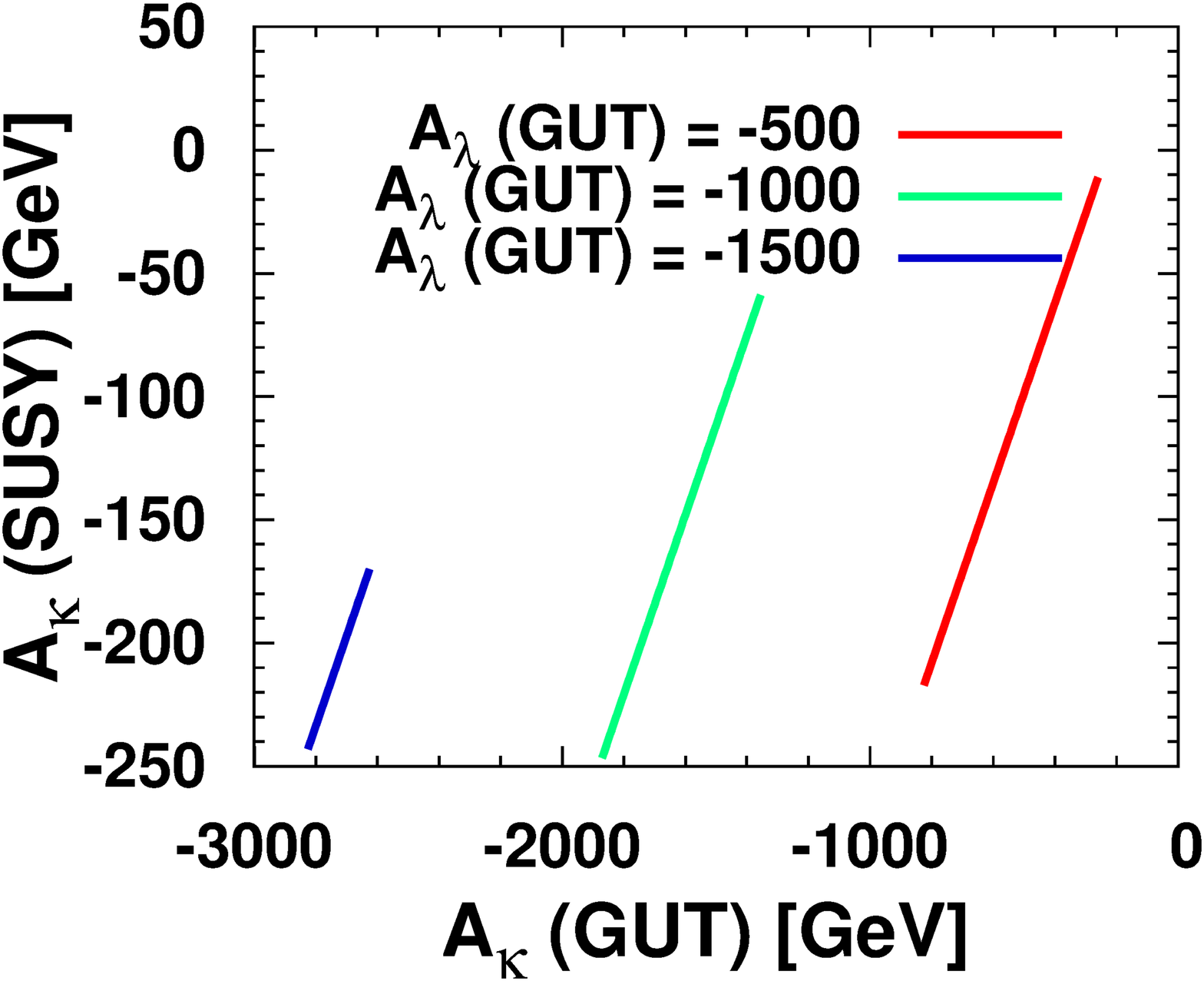} 
 \end{center}
 \caption{ 
Left: the mass of the lightest Higgs boson $m_{H_1}$ can become rather low for large negative values of  $A_{\kappa}$. However, in 
this case the annihilation channels $\chi \chi \rightarrow H_1 H_1$ or $\chi \chi \rightarrow Z H_1$ become 
kinematically allowed, which leads to a too small value for the relic density, as indicated by the color coding. For 
values of $A_{\kappa}$ near zero the value of the relic density drops again because the annihilation into two light 
pseudo-scalar Higgs bosons becomes allowed. 
Right:  large negative values of  $A_{\kappa}$ -- implying small $m_{H_1}$ -- can only be reached for 
large negative values of $A_{\lambda}$ and $A_{\kappa}$   at the GUT scale. The different lines are drawn for different 
values of $A_\lambda$. They 
 show the strong correlations between $A_\kappa$(SUSY), $A_\kappa$(GUT) and $A_\lambda$(GUT).
 Such strong correlations require a careful 
fitting method with all parameters left free, as discussed in Sect. \ref{multi}.}\label{f6}
 \end{figure}

\subsection{Limits on Higgs boson masses} 
Masses of the heaviest Higgs boson depend on the chosen trilinear couplings, see Table \ref{t1} and Table 
\ref{t3},\ref{t4}. Therefore no upper limit on $m_{H_3}$ can be found, if trilinear couplings in the multi-TeV range are 
allowed. Only a lower limit of around 280 GeV for $m_{H_3}$ can be obtained from the lower limit on the chargino mass, 
which provides a lower limit on  $\mu_{eff}$.   For  scenarios, where the second lightest Higgs boson has a mass of  126 
GeV, one observes the lightest Higgs boson to be in the range of a few GeV to 126 GeV. However, if one assumes the relic 
density is provided by the LSP, the lower limit increases to about 60 GeV, since else the relic density becomes too 
small,  as can be seen from  the color code in the left plot in Fig. \ref{f6}. For $H_1$ masses below about 60 GeV
the LSP annihilation into two lightest Higgs bosons and/or a Higgs and a $Z^0$ boson becomes kinematically allowed, thus 
leading to a too low relic density. Fig. \ref{f6} demonstrates also, that low values of $m_{H_1}$ can only be obtained 
for moderate negative values of
$A_\kappa$ at the SUSY scale, which can only be obtained for large negative   values of $A_\kappa$ and $A_\lambda$
at the GUT scale, as demonstrated in the right panel of Fig. \ref{f6}.

\section{Summary}
In this Letter we have compared the CMSSM with the NMSSM. 
A Higgs boson of 126 GeV requires stop masses well above 3 TeV in the CMSSM, if the trilinear couplings are defined at 
the GUT scale, which leads to large radiative corrections with fixed point solutions at the electroweak scale. Such 
fixed point solutions do not allow maximal mixing in the stop sector independent of the GUT scale scenario, which leads 
to the multi-TeV stop masses. 
In the NMSSM the Higgs mass gets corrections at tree level from the mixing with the additional singlet, so the stop 
masses are not required to be heavy to provide large radiative corrections.

We use  NMSSMTools to compare data with theory, taking into account the full 1-loop corrections in the NMSSM with 
additional 2-loop corrections \cite{Degrassi:2009yq}. These corrections lower the Higgs boson mass by several GeV, so it 
is important to include them. In practice, it means that one cannot reach the 126 GeV Higgs boson mass anymore with the 
CMSSM-like portion of the NMSSM parameter space, i.e. small couplings of $\lambda$ and $\kappa$.
Instead, one has to resort to the full 9-dimensional parameter space of the NMSSM, which becomes challenging.
Using the multi-step fitting technique of Paper I the strong correlations between the parameters are taken into account 
and the full parameter space is sampled efficiently.

The additional NMSSM couplings in the Higgs sector are rather well constrained by the observed Higgs boson mass of 126 
GeV in combination with the relic density (Fig. \ref{f2}), which determine the ratio of vevs of the Higgs doublets and 
lead to large  values of $\tb$ in the CMSSM ($\approx 50$ in most of the parameter space) and small values of $\tb$ in 
the order of a few in the NMSSM  (Fig. \ref{f3}). One of the two lightest neutral scalar Higgs bosons of the NMSSM is 
SM-like, while the other one is singlet-like with small reduced couplings, so deviations of the couplings from the SM 
for the 126 GeV are easily obtained by changing the mixing and hence  the mass splitting between the two lightest Higgs 
bosons.
 
We find that the excluded SUSY mass parameter region in the $\mzero,\mhalf$ plane is dominated by the Higgs boson mass 
for the CMSSM because of the required multi-TeV stop masses, while in the NMSSM the excluded region is determined by the 
LHC SUSY searches only (Fig. \ref{f4}). The dark matter constraints in the NMSSM are easily fulfilled because of the 
large Higgsino component of the LSP, which can be changed by varying the mixing in the Higgs sector.

The large triple Higgs couplings in the NMSSM  lead to very different Higgs boson decays: in the CMSSM  the heavier 
Higgs boson couples preferentially to d-type fermions, while in the NMSSM the dominant decay modes are the lighter Higgs 
bosons and gauginos with a significant Higgsino component. So the  unique NMSSM search signatures at the LHC are the 
double Higgs boson production, i.e. two Higgs bosons in a single event, one of them having a mass of 126 GeV and  a 
second Higgs boson below or above 126 GeV. In addition, since the LSP is higgsino-like, Higgs boson decays into LSPs can 
be appreciable, thus leading to invisible Higgs decays.\\[0.2cm]
\section*{Acknowledgements}
Support from the Deutsche Forschungsgemeinschaft (DFG) via a Mercator Professorship (Prof. Kazakov) is warmly 
acknowledged.
We thank U. Ellwanger for helpful discussions regarding NMSSMTools. 






\bibliographystyle{lucas_unsrt}
\bibliography{nmssm_paper_short}


\end{document}